\documentclass{JHEP3}
\usepackage{amsmath}
\usepackage{epsfig}
\usepackage{amssymb}
\usepackage{graphics}
\usepackage{url}

\newcommand{\lba}{\left[\begin{array}}
	\newcommand{\rba}{\end{array}\right]}
\newcommand{\bea}{\begin{eqnarray}}
	\newcommand{\eea}{\end{eqnarray}}
\newcommand{\lam}{\lambda}

\newcommand{\bean}{\begin{eqnarray}}
	\newcommand{\eean}{\end{eqnarray}}
\newcommand{\nn}{\nonumber \\}

\def\W #1{\widetilde{#1}}

\def\eref#1{(\ref{#1})}

\def\d{{\rm d}}

\def\a{{\alpha}}

\def\c{{\gamma}}

\def\d{\partial}

\def\la{\lambda}

\def\Label#1{\label{#1}%
	\smash{\hbox to0pt{\raise1ex\hbox{\tiny[#1]}\hss}}}

\title{Reduction of two-loop Feynman integrals in parametric representation with syzygy trick }
\author{Hongbin Wang \footnote{Emails: 21836003@zju.edu.cn } \\
	{\small Zhejiang Institute of Modern Physics, Zhejiang University, Hangzhou, 310027, P. R. China }
}
\date{\today}
\abstract{Reduction  of high-loop Feynman integrals is one of the main tasks in scatting amplitude. In this paper, a new representation of Feynman integrals proposed by Chen in \cite{chen1,chen2} is considered.  We combined Chen's method with "syzygy" trick to simplify the IBP relations, and successfully canceled the dimensional shift and the unwanted doubled propagators. Moreover, we improved the method to deal with tensor's structure. We demonstrated our method using 
 three two-loop integrals to show our method, and presented the analytical reduction coefficients in the top-sector.
}


\newpage
\begin{document}
	\section{Introduction}
	The analytic calculation of scatting amplitude for a given process is an interesting and important task. Since the complexity for high-loop calculation, many new and novel technique have been proposed in recent years. The most used method is the Integrating-by-Part(IBP) method \cite{smir,ibp1,ibp2}, which reduce the Feynman integrals to the master integrals(MI). However, to get the reduction results, there are usually a great number of IBP relations in practical calculations, which is hard to solve. Finding more efficient reduction methods becomes an important direction.
	%
	%
	One way to  calculate and reduce the Feynman integrals is to establish  representation in other space rather than the traditional momentum space. There are many well-known representation such as Feynman parametric representation \cite{Bern:1992em,Bern:1993kr} and  Baikov representation \cite{Baikov:1996rk,Baikov:1996cd}.
	
	In recent years, Chen proposed a new parametric representation for Feynman integrals \cite{chen1,chen2}. One of the benifit in the new representation is that  the number of IBP identities is usually less than he traditional momentum representation. For example, with $n$  propagators  the number of IBP relations in Chen's form is
	 $n+1$. However, to find the recursion relation using the IBP identities in the parametric representation, there will naturally be some terms in different spacetime dimensions. Since we usually care about the reduction in the same dimension $D$, we need to cancel the dimensional shift. The trick  to cancel the unwanted terms is known as solving "syzygy" equation \cite{Kosower,zhang1,zhang2,Larsen:2015ped,Larsen:2016tdk,Zhang:2016kfo,Jiang:2017phk}.

	In our recent work \cite{2021General}, we consider the Chen's method \cite{chen1,chen2} at one-loop case, when the homogeneous polynomial $F$ is at degree 2. By this fact  we proposed a method by adding an antisymmetry matrix $\hat K_{A}$ with many parameters. By choosing the parameters, we are able to cancel the  terms with dimension shift in the IBP relation, and at the same time simplified the IBP relations by dropping the integrals with doubled propagators. In this paper, we considered the two-loop case, when the polynomial $F$ is at degree 3. We proposed the syzygy trick in the parametric representation and gave some explicit examples. By the technique, we got the independent IBP relations which does not contain the dimensional shift and the unwanted terms with  higher power for propagators. Further more, we considered the tensor's case. After giving a brief review of the tensor's case in Chen's parametric representation, we gave some examples to show the "syzygy" trick  in reduction of tensor's integrals.

	The plan of this paper is following. In section 2, we gave a brief review of Chen's parametric representation of Feynman integrals in scalar's and tensor's cases. In section 3, we show how we improved the IBP recurrence relation to cancel the dimensional shift and simplify the IBP relations  in both scalar's and tensor's case. By the improved IBP identities, in  section 4, 5, and 6, we gave three different examples to show our method and compared it with others. Since the analytic expressions for various quantities are long, some of them have been presented in the Appendix.
\section{Review of Chen's method}

In this section, we will briefly discuss the
method proposed in \cite{chen1,chen2}
by Chen. This method depends crucially on a
different parametrization form, which is obtained by
adding one more integral variable over the
familiar Feynman parametrization.   For example, for the
scalar integrals
\bea
I(L;\lambda_1+1,\cdots ,\lambda_n+1)&=&\int d^Dl_1\cdots d^Dl_L\frac{1}{D_1^{\lambda_1+1}\cdots D_{n}^{\lambda_n+1}}~~~\label{Wang-2-2}
\eea
the familiar Feynman parametrization leads to
\bea
I(L;\lambda_1+1,\cdots ,\lambda_n+1)&=&(-i)^{n+ \lambda-\frac{DL}{2}}\Gamma(n+\lambda-\frac{DL}{2})\int dx_1\cdots dx_n\delta (\sum_{j\in S} x_j-1)U^{\lambda_u}f^{\lambda_f}x_1^{\lambda_1}\cdots x_n^{\lambda_n}~~~\label{1.18}
\eea
where
\bea
U(x) & = & Det~A,~~~~~~~~f(x)=-V(x)+U(x)\sum m_i^2x_i\nn
\lambda&=&\sum_{i=1}^{n} \lambda_i,~~~~
\lambda_u=n+\lambda-\frac{D}{2}(L+1),~~~~
\lambda_f=-n-\lambda+\frac{DL}{2}~~~\label{Wang-2-8}
\eea
and $U(x),V(x)$ are defined through
\bea
& & \sum_{i}^{L}\a_iD_i=\sum_{i,j}^{L}A_{ij}l_i\cdot l_j+2\sum_{i=1}^{L}B_i\cdot l_i+C\nn
& & C-\sum A_{ij}^{-1}B_i\cdot B_j\equiv \frac{V(\a)}{U(\a)}-\sum m_i^2\a_i ~~~\label{Wang-2-5}
\eea
To go to Chen's parametrization, we use the Mellin transformation
\bea
A^{\lambda_1}B^{\lambda_2}&=&\frac{\Gamma(-\lambda_1-\lambda_2)}{\Gamma(-\lambda_1)\Gamma(-\lambda_2)}\int _0^{\infty} dx  (A+Bx)^{\lambda_1+\lambda_2}x^{-\lambda_2-1}~~~\label{Wang-2-9}
\eea
to rewrite \eref{1.18} as
\cite{chen1,chen2} \footnote{Notice that the $\lambda_i$ is different from the power in the integral in \eref{Wang-2-2}.}
\bea
I[\lambda_1+1,\cdots ,\lambda_n+1]&=&(-1)^{n+\lam} i^{L}\pi^{\frac{LD}{2}}\frac{\Gamma(-\lambda_0)}{\Pi_{i=1}^{n+1}\Gamma(\lambda_i+1)}i[\lambda_0;\lambda_1,\cdots ,\lambda_n]~~~~\Label{defI}
\eea
with \footnote{Here the F is given by $F=U(B^T A^{-1}B -C +x_{n+1})$, with the $A$, $B$ and $C$ are given in \eref{Wang-2-5}
	.}
\bea
&&\lambda_0=-\frac{D}{2},~~~\lam\equiv \sum_{i=1}^{n}\lambda_i,~~\lambda_{n+1}=-n-1-\sum_{i=1}^{n}\lambda_i+\frac{LD}{2},~~~F=U x_{n+1}+f\nn
&&i[\lambda_0;\lambda_1,\cdots ,\lambda_n]\equiv \int d\Pi^{(n+1)}F^{\lambda_0}x_1^{\lambda_1}\cdots x_n^{\lambda_n}x_{n+1}^{\lambda_{n+1}},~~~d\Pi^{(n+1)}\equiv dx_1\cdots dx_{n+1}\delta(\sum_{i\in S}x_i-1)~~~~\Label{pin+1}
\eea
%
%
Here $S$ is an arbitrary non-trivial subset of $\{1,2,\cdots n\}$.
Notice the $\lambda_0$ is the dimensional regularization parameter with the space-time dimension $D$.	

To deal with tensor integrals
\bea I[\lambda_0;\lambda_1+1,\cdots,\lambda_n+1]^{\mu_1\mu_2\cdots \mu_m}&\equiv & \int d^{D}l_1\cdots d^{D}l_L \frac{l_{i_1}^{\mu_1}\cdots l_{i_m}^{\mu_m}}{D_1^{\lambda_1+1}\cdots D_n^{\lambda_n+1}}~~~~~~\Label{tendef}
\eea
we could  use the identity
\bea
l_{i_1}^{\mu_1}l_{i_2}^{\mu_2}\cdots l_{i_m}^{\mu_m}&=&\frac{i(-1)^m}{\Gamma(m+1)}\Big\{\frac{\partial }{\partial q_{i_1,\mu_1}}\frac{\partial }{\partial q_{i_2,\mu_2}}\cdots \frac{\partial }{\partial q_{i_m,\mu_m}}\int _0^{\infty}dy \exp[-iy(1+\sum_{i=1}^{L}q_i\cdot l_i)]\Big\}|_{q_i^\mu\to0}
\eea
to rewrite  the \eref{tendef} as
\bea
&&I[\lambda_0;\lambda_1+1,\cdots\lambda_n+1]^{\mu_1\mu_2\cdots \mu_m}\nn
&=&\frac{i(-1)^{m}}{\Gamma(m+1)}\frac{\d }{\d q_{i_1,\mu_1}}\cdots\frac{\d }{\d q_{i_m,\mu_m}}\int \frac{d^{D}l_1\cdots d^{D}l_L}{D_1^{\lambda_1+1}\cdots D_n^{\lambda_{n}+1}}\int _0^{\infty}dy \exp[-iy(1+\sum_{i=1}^Lq_i\cdot l_i)]|_{q_i^{\mu}\to0}~~~~\Label{tensor1}
\eea
The integration of $y$ can be carried out to give ${1\over D_0}$ with $D_0=1+\sum_{i=1}^m q_i\cdot l_i$. 
In other words, the $\int dy$ can be interpreted as the Schwinger parametrization of ${1\over D_0}$. 
Doing the Schwinger parametrization for other $D_i, i=1,...,n$, it becomes the scalar integrals discussed in 
\eref{Wang-2-2} and we have 
\bea
\sum_{i=1}^{n}x_iD_i-y D_0&\equiv &\sum_{i,j=1}^L A_{ij}l_i\cdot l_j+2\sum_{i=1}^L \W B_i \cdot l_i+\W C,~~~~\W B_i=B_i-\frac{y}{2}q_i,~~~~\W C=C-y
\eea
for which we can read out
\bea
U(x)=Det[A],~~~f(q,y)&=&U(\W B^T A^{-1}\W B -\W C),~~F(q,y)=x_{n+1}U+f(q,y)
\eea
After doing the integration over $l_1$ to $l_L$
we arrive
\bea
I[\lambda_0;\lambda_1+1,\cdots ,\lambda_n+1]^{\mu_1\cdots \mu_m}&=&(-1)^{n+\lam+1}i^{L}\pi^{\frac{LD}{2}}\frac{\Gamma(-\lambda_0)}{\Pi_{i=1}^{n}\Gamma(\lambda_i+1)\Gamma(\lambda_{n+1})}i[\lambda_0;\lambda_1,\cdots ,\lambda_n]^{\mu_1\cdots\mu_m}~~~\Label{Itoi}
\eea
where
\bea
&&i[\lambda_0;\lambda_1,\cdots,\lambda_n]^{\mu_1\cdots\mu_m}\nn
&\equiv &\frac{(-1)^m}{\Gamma(m+1)}\frac{\d }{\d q_{i_1,\mu_1}}\frac{\d }{\d q_{i_2,\mu_2}}\cdots\frac{\d }{\d q_{i_m,\mu_m}}\int dx_1\cdots dx_{n+1}dy\delta(\sum_{j\in S}x_j-1)F(q,y)^{\lambda_0}x_1^{\lambda_1}\cdots x_{n+1}^{\lambda_{n+1}-1}|_{q_i\to0}\nn
&=&\frac{(-1)^{m}}{\Gamma(m+1)}\frac{\d }{\d q_{i_1,\mu_1}}\cdots\frac{\d }{\d q_{i_m,\mu_m}}\int d\Pi^{(n+1)}\int _0^{\infty }dy~F(q,y)^{\lambda_0}x_1^{\lambda_1}\cdots x_{n+1}^{\lambda_{n+1}-1}|_{q_i\to0}~~~\Label{tensor2}
\eea
It is worth noticing that  here we have totally $n+1$ propagators, so there is a factor $\frac{1}{\Gamma(\lambda_{n+1})}$ not $\frac{1}{\Gamma(\lambda_{n+1}+1)}$ which fixing the power of $x_{n+1}$ with the definition of $\lambda_{n+1}$ as  $\lambda_{n+1}\equiv -(n+1)-\sum_{i=1}^{n}\lambda_i+\frac{LD}{2}$.
To continue,  we  decompose the function $F(q,y)$ into the form
\bea
F(q,y)&=&F(0,0)+yU-yU\sum_{i,j=1}^L[A^{-1}_{ij} B_i\cdot q_j]|_{y=0}+\frac{1}{4}y^2U\sum_{i,j=1}^L[A^{-1}_{ij}q_i\cdot q_j]|_{y=0}\nn
&=&F(0,0)+yU+y\sum_{=1}^L b_i \cdot q_i+\sum_{i,j}^L c_{ij}y^2 q_i\cdot q_j~~~\label{q-dep}
\eea
where $F(0,0)$ is the same one in \eref{pin+1} and  $b$ and $c$ are  polynomials of x's
\bea
b_i&=&-\sum_{i=1}^L -UA_{ij}^{-1}\W B_j|_{y=0}=-\sum_{i=1}^L U A_{ij}^{-1} B_j,~~~~~~~c_{ij}=\frac{1}{4}UA_{ij}^{-1}|_{y=0}=\frac{1}{4}U A_{ij}^{-1}~~~\Label{b,c}
\eea
Since the $q$ dependence has been explicitly written done in \eref{q-dep},  carrying out
the differential operators $\frac{\d }{\d q_{i_j,\mu_j}}$ in \eref{tensor2} we arrive following sum
\bea
\sum_{n_0=\la_0-m}^{\la_0}\sum_{ i_t}
c[n_0]^{\mu_1\cdots\mu_m}_{i_1,\cdots, i_n}  \int d\Pi^{(n+1)}\int _0^{\infty}dy \Big\{F(0,0)+yU\Big\}^{n_0}y^{\delta_y}x_1^{\lambda_1+i_1}\cdots x_n^{\lambda_n+i_n}x_{n+1}^{\lam{n+1}-1}~~~~~~~~~~~~~\Label{three}
\eea
where the sum over all possible choices of $i_t$ with $t=1,...,n$ such that 
\bea i_t\geq 0,~~~~~\sum_{t=1}^n i_t=(\lambda_0-n_0)(1+L)-m\eea
The coefficients $c[n_0]^{\mu_1\cdots\mu_m}_{i_1,\cdots, i_n}$ can be worked out although  expressions
are little complicated. Another point is that although the sum is up to $\la_0$, one can check that 
with $n_0>\frac{2\lambda_0-m}{2}$ $c[n_0]$  is always zero by the degree of $q_i$s.


Recalling the $F(0,0)$ given in \eref{pin+1}, we can combine $x_{n+1}+y=\W x_{n+1}$ or do the replacement
\bea
x_{n+1}\to x_{n+1}-y
\eea
and the \eref{three} becomes to
\bea
\frac{(-1)^m}{\Gamma(m+1)}\sum_{\substack{n_0=\lambda_0-m\\i_1+\cdots i_n+m\\=(\lambda_0-n_0)(L+1)}}^{\lambda_0} && c[n_0]^{\mu_1\cdots\mu_m}_{i_1,\cdots, i_n}\int d\Pi^{(n+1)}\int dy\Big\{F(0,0)\Big\}^{n_0}\times y^{\delta_y}x_1^{\lambda_1+i_1}\cdots x_n^{\lambda_n+i_n}(x_{n+1}-y)^{\lambda_{n+1}-1}~~~\Label{tensor4}~~~
\eea
Using
\bea
\int  dy y^{\delta_y}(x_{n+1}-y)^{\lambda_{n+1}-1}=\frac{x_{n+1}^{\lambda_{n+1}+\delta_y}\Gamma(1+\delta_y)\Gamma(\lambda_{n+1})}{\Gamma(1+\delta_y+\lambda_{n+1})}
\eea
we got
\bea
&&\frac{(-1)^m}{\Gamma(m+1)}\sum_{\substack{n_0=\lambda_0-m\\i_1+\cdots i_n+m\\=(\lambda_0-n_0)(L+1)}}^{\lambda_0}  c[n_0]^{\mu_1\cdots\mu_m}_{i_1,\cdots, i_n}\frac{\Gamma(1+\delta_y)\Gamma(\lambda_{n+1})}{\Gamma(1+\delta_y+\lambda_{n+1})}\int d\Pi^{(n+1)}F(0,0)^{n_0}x_1^{\lambda_1+i_1}\cdots x_n^{\lambda_n+i_n}x_{n+1}^{\lambda_{n+1}+\delta_y}\nn
&=&\frac{(-1)^m}{\Gamma(m+1)}\sum_{\substack{n_0=\lambda_0-m\\i_1+\cdots i_n+m\\=(\lambda_0-n_0)(L+1)}}^{\lambda_0} c[n_0]^{\mu_1\cdots\mu_m}_{i_1,\cdots, i_n}\frac{\Gamma(1+\delta_y)\Gamma(\lambda_{n+1})}{\Gamma(1+\delta_y+\lambda_{n+1})}i[n_0;\lambda_1+i_1,\cdots,\lambda_n+i_n]~~~\Label{sum}
\eea
Putting the result \eref{sum} back to \eref{Itoi}, we have
\bea
I[\lambda_0,\lambda_1+1,\cdots,\lambda_{n}+1]^{\mu_1\cdots\mu_m}&= & (-1)^{n+\lam+m}i^{L+1}\pi^{\frac{LD}{2}}\frac{\Gamma(-\lambda_0)}{\Pi_{i=1}^n\Gamma(\lambda_i+1)}\nn & & \times\sum_{\substack{n_0=\lambda_0-m\\i_1+\cdots i_n+m\\=(\lambda_0-n_0)(L+1)}}^{\lambda_0} c[n_0]^{\mu_1\cdots\mu_m}_{i_1,\cdots, i_n} \frac{1}{\Gamma(m+\lambda_{n+1}+1)}i[n_0;\lambda_1+i_1,\cdots,\lambda_n+i_n]~~~~~~~~~~
\Label{sumI}
\eea
where we have used the fact that  after  taking the differential operators and setting the limit of $q_i\to0$,   $\delta_y=m$.
%

To demonstrate the usage of \eref{sumI}, let us present two examples for tensor one-loop bubbles. Let us start
with the tensor rank one, i.e.,
%
\bea
I_2(n_1+1,n_2+1)^{\mu}&=&\int d^Dl\frac{l^{\mu}}{(l^2-m_1^2)^{n_1+1}((l-p)^2-m_2^2)^{n_2+1}}\nn
&=&(-1)^{2+n_1+n_2}i\pi^{\frac{D}{2}}\frac{\Gamma(-\lambda_0)}{\Gamma(n_1+1)\Gamma(n_2+1)\Gamma(\lambda_3)}i[\lambda_0;n_1,n_2]^{\mu}~~~\Label{I2111}
\eea
where 
\bea
i[\lambda_0;n_1,n_2]^{\mu}&=&\frac{-1}{\Gamma(2)}\frac{\d }{\d q_{1,\mu}}\int d\Pi^{(3)}\int_0^{\infty}dy F(q,y)^{\lambda_0}x_1^{n_1}x_2^{n_2}x_3^{\lambda_3-1}|_{q_1\to 0}~~~\Label{bubblen1n2}
\eea
with the polynomial $F(q,y)$
\bea
F(q,y)&=&-p^2 x_1 x_2 + m_1^2 x_1 (x_1 + x_2) +
m_2^2 x_2 (x_1 + x_2) + (x_1 + x_2) x_3 + (x_1 + x_2 + p\cdot  q_1 x_2) y + \frac{q_1^2y^2}{4}~~~~~~~
\eea
Carrying out  $\frac{\d }{\d q_{1,\mu}}$ the \eref{bubblen1n2} becomes to
\bea
i[\lambda_0;n_1,n_2]^{\mu}&=&\frac{-\Gamma(\lambda_3)p^\mu \lam_0}{\Gamma(2+\lambda_3)}i[\lambda_0-1;n_1,n_2+1]~~~\Label{exab}
\eea
with $c[\lambda_0-1]^{\mu}_{0,1}=-\lam_0 p^{\mu}$, 
thus 
\bea
I_2[n_1+1,n_2+1]^{\mu}&=&(-1)^{n_1+n_2}\pi^{\frac{D}{2}}\frac{\Gamma(-\lambda_0)\lam_0p^\mu}{\Gamma(n_1+1)\Gamma(n_2+1)\Gamma(2+\lambda_3)}i[\lambda_0-1;n_1,n_2+1]
\eea
%
%
For bubbles with tensor rank two, by  \eref{tensor1} we have  
\bea
I_2[n_1+1,n_2+1]^{\mu\nu}&=&(-1)^{2+n_1+n_2}i \pi^{\frac{D}{2}}\frac{\Gamma(-\lam_0)}{\Gamma(n_1)\Gamma(n_2)\Gamma(\lam_3)}i[n_1,n_2]^{\mu\nu}
\eea
where (by \eref{tensor2})
\bea
i[n_1,n_2]^{\mu\nu}&=&\frac{1}{\Gamma(3)}\frac{\d }{\d q_{1,\mu}}\frac{\d }{\d q_{1,\nu}}\int d\Pi^{(3)}\int _0^{\infty} dy F(q,y)^{\lam_0}x_1^{n_1}x_2^{n_2}x_3^{\lam_3-1}|_{q_1\to0},~~~~\lam_3=-3-n_1-n_2-2\lam_0~~~~~~~~~
\eea
wtih  the corresponding function 
\bea
F(q,y)&=&-p^2 x_1 x_2 + m_1^2 x_1 (x_1 + x_2) +
m_2^2 x_2 (x_1 + x_2) + (x_1 + x_2) x_3 + (x_1 + x_2 + p\cdot  q_1 x_2) y + \frac{q_1^2y^2}{4}~~~~~~~~~~
\eea
Carrying out the derivative, we get two terms
\bea
&&i[n_1,n_2]^{\mu\nu}=\lam_0(\lam_0-1)p^\mu p^\nu \frac{\Gamma(\lam_3)}{\Gamma(\lam_3+3) }i[\lam_0-2;n_1,n_2+2]+\frac{1}{2}\lam_0 g^{\mu\nu}\frac{\Gamma(\lam_3)}{\Gamma(\lam_3+3) } i[\lam_0-1;n_1,n_2]~~~\Label{tensorfram2}
\eea
where two  coefficients are 
\bea
c[\lam_0-2]^{\mu\nu}_{0,2}&=& \lam_0(\lam_0-1) p^\mu p^\nu,~~~c[\lam_0-1]^{\mu\nu}_{0,0}=\frac{1}{2}\lam_0 g^{\mu\nu}
\eea
with $\lam_0=-\frac{D}{2}$.
\section{The improved IBP relations with syzygy method}
%
Having reduced everything to the integral form \eref{pin+1}, we can establish the corresponding IBP relations in this frame. It is given by  \footnote{Notice that in the first term, the power of $x_{n+1}$ has been shifted before the differential, which keeps the degree of the integrand  $-n-1$. In general, one can choose arbitary $x_i$ to shift the degree, but for simplicity, we chose to shift the $x_{n+1}$ here.}
\bea
\int d\Pi^{(n+1)}\frac{\partial}{\partial x_i}\Big\{F^{\lambda_0}x_1^{\lambda_1}\cdots x_n^{\lambda_n}x_{n+1}^{\lambda_{n+1}+1}\Big\}+\delta_{\lambda_i,0}\int d\Pi^{(n)}\Big\{F^{\lambda_0}x_1^{\lambda_1}\cdots x_n^{\lambda_n}x_{n+1}^{\lambda_{n+1}+1}\Big\}|_{x_i\rightarrow 0}&=&0,~~i=1,\cdots n+1~~~~~~~\Label{ibpiden}
\eea
where the second term is the boundary term  which contributes to the sub-topology, i.e. the integral with the propagator $D_i$ having been removed. Notice in the first term the sum of the power of $x_i$ is one more than that in $i[\lambda_0;\lambda_1,\cdots , \lambda_n]$. There are total $(n+1)$ independent IBP recurrence relations. 

Since the action of differential operator $\frac{\partial }{\partial x_i}$  over the function $F$ will produce the integrals in different dimension, which makes the reduction procedure cumbersome,   we will use the syzygy trick suggested in [cite Kosower, Zhang etc] \footnote{In \cite{Needed}, Zhang has proposed the method in Baikov represent.}.(1,2,7,8,9,10,11,12,13)

To overcome the difficulties and simplify the reduction procedure.

%

Let us combining the $n+1$ identities \eref{ibpiden}, with each one multiplied a polynomial factor $z_i$ in degree zero\footnote{Here we mean the  term like  $z_1=\frac{x_{i_1}^{\c_1}x_{i_2}^{\c_2}\cdots x_{i_k}^{\a_k}}{x_{n+1}^{\c_1+\c_2+\cdots+\a_k}}$, for the reason that the integrand of \eref{IBPsyz} must be degree $-n-1$, the factor $z_i$ must keep the degree of the integrand.}
\bea
\int d\Pi^{(n+1)} \sum_{i=1}^{n+1}\frac{\d }{\d x_i}\Big\{z_iF^{\lambda_0}x_1^{\lambda_1}\cdots x_n^{\lambda_n}x_{n+1}^{\lambda_{n+1}+1}\Big\}+\delta_{R}&=&0~~~\Label{IBPsyz}
\eea
Expanding \eref{IBPsyz} we got
\bea
\int d\Pi^{(n+1)} \Big\{\sum_{i=1}^{n+1} \frac{\d z_i}{\d x_i}+\lambda_0 \frac{\sum_{i=1}^{n+1}z_i\frac{\d F}{\d x_i}}{F}+\sum_{i=1}^{n+1}\frac{\lambda_iz_i}{x_i}+\frac{z_{n+1}}{x_{n+1}}\Big\}F^{\lambda_0}x_1^{\lambda_1}\cdots x_n^{\lambda_n}x_{n+1}^{\lambda_{n+1}+1}+\delta_{R}&=&0~~~\Label{syz2}
\eea
with the combined  boundary term $\delta_R$ \footnote{Actually in the general case, the parameter $\lambda_{n+1}$ will never be zero, so the summary could be written as $\sum_{i=1}^n$ and drop the case when $i=n+1$.}
\bea
\delta_R&=&\sum_{i=1}^{n+1}\int d\Pi^{(n)}\Big\{z_iF^{\lambda_0}x_1^{\lambda_1}x_2^{\lambda_2}\cdots x_n^{\lambda_n}x_{n+1}^{\lambda_{n+1}+1}\Big\}|_{x_i\to0}
\eea
%
Since the second term in \eref{syz2} contains a $F$ in denominator, it will shift the spacetime dimension by two (recalling that  $\lambda_0=-\frac{D}{2}$). In practical calculation  we usually need the IBP relation without dimensional shift. 
To avoid this, we  use the trick of "syzygy" in computational algebranic geometry \cite{zhang1,zhang2,Larsen:2015ped,Larsen:2016tdk,Zhang:2016kfo,Jiang:2017phk}, i.e., choosing the functions $z_i$. 
such that\footnote{Since the degree of both side is equal, the solution of $B$ is a polynomial of degree $-1$.}
\bea
\sum_{i=1}^{n+1}z_i\frac{\d F}{\d x_i}&=&BF~~~~\Label{2.5}
\eea
In general, solving the syzygy of the type \eref{2.5} is highly nontrivial. However, 
since the $F(x)$ is a homogeneous function of $x_1$ to $x_{n+1}$ of degree $L+1$, there is a natural and trivial solution of $z_i$ by the Euler equation for homogeneous function,
\bea
\sum_{i=1}^{n+1} x_i\frac{\d F}{\d x_i}&=&(L+1)F ~~~\Label{eu}
\eea
To keep the degree, we multiply a $\frac{1}{x_{n+1}}$ in both side of \eref{eu}\footnote{The choice of denominator is free from $x_1$ to $x_{n+1}$, but the most natural choice is just $x_{n+1}$.}, i.e., the trivial solution $\W z_i=\frac{x_i}{x_{n+1}}$ such that 
\bea
\sum_{i=1}^{n+1}\W z_i \frac{\d F}{\d x_i}&=&\frac{L+1}{x_{n+1}}F~~~\label{2.7}
\eea
With above trivial solution, we can split a general solution $z_i$ as 
%
\bea
z_i\to z_i+{B x_{n+1}\over L+1}\W z_i~~~\Label{decom}
\eea
then syzygy can be split to
\bea
\sum_{i=1}^{n+1} z_i\frac{\d F}{\d x_i}&=&0,~~~~~~\sum_{i=1}^{n+1} \tilde z_i \frac{\d F}{\d x_i}=BF~~~\Label{2z}
\eea
%
Since the $\tilde z_i$ part always give the trivial IBP relation $0=0$,
to find useful IBP reduction relations we just need to find the solution for the 
first  equation in \eref{2z}, which could be solved by SINGULAR
in the form of polynomials,
\bea
\sum_{i=1}^{n+1} g_{ji}\frac{\d F}{\d x_i}&=&0,~~~j=1\cdots s_{k}~~~\Label{syzy}
\eea
The solution is called the syzygy of the polynomial ring of $F[\frac{\d F}{\d x_1},\cdots,\frac{\d F}{\d x_{n+1}}]$,
where $s_k$ is the number of generators to the syzygy $g_i$. The $g_{ji}$s are all homogeneous function of $x_1$ to $x_{n+1}$.
For two-loop integrals, by our experiences there are usually 10 to 100  independent syzygy generators.
Then the $z_i$ could be chosen as
\bea
z_i&=&\sum_{j=1}^{s_k}c_j{g_{ji}}{x_{n+1}^{-k_j}}~~~~\Label{2.11}
\eea
where the addtional factor $x_{n+1}^{-k_j}$ is to make the $z_i$ degree zero.
The $c_j$ could be degree zero rational function of $x_i$, but for simplicity, one can choose them to be independent of $x_i$ and be the functions of  kinematic variables $s,t,m_i$ only. 
Even with such a simplification, there are still various choices for $c_i$. As we will show, with
some nice choices of $c_j$s, one can get  IBP realtions,  where there is no  terms with increasing power of propagators.
This will reduce the complexity of reduction a lot.

%
%
%
When we consider the tensor reduction, there are some differences. First from 
\eref{sumI}, one can see that for the tensor rank $m$, the values of $n_0$ can be from
$\lambda_0-m$ to $\lambda_0$. In the case $n_0\neq \lambda_0$, the dimension has been shifted. To rewrite the integrals in different shifted dimension into the original dimension $\lam_0=-\frac{D}{2}$, we need to get the IBP relation connecting different dimensions in \eref{syz2}. Now the syzygy \eref{syzy} could be modified to
%
\bea
\sum_{i=1}^{n+1}g_{ji}\frac{\d F}{\d x_i}&=&g_{j0}x_1^{\c_1}x_2^{\c_2}\cdots x_{n}^{\c_n}x_{n+1}^{\c_{n+1}}     ~~~~\Label{tensortrick1}
\eea
where the role of $x_i^{\gamma_i}$ is to shift the power of propagators to the wanted number. 
Putting it to \eref{syzy} we will get a recurrence  relation where  there is one term 
\bea
\lam_0 g_{j0}\int d\Pi^{(n+1)} F^{\lam_0-1}x_1^{\lam_1+\gamma_1}\cdots x_{n}^{\lam_n+\gamma_n}x_{n+1}^{\lam_{n+1}+\gamma_{n+1}}.
\eea
This term corresponds to dimension $(D+2)$ while other terms, dimension $D$. 
Now we successfully got a recurrence relation of integrals between dimension $D$ and $D+2$ (or in general, $D+k+2$ to $D+k$). Repeating this procedure, we will finally reduce integrals in dimension parameter $n_0$ in \eref{sumI} from $n_0=\lam_0-m$ to $\lam_0$.
%
%

%
%
%
%
%
\section{Example 1: massive sunset with different mass}

Having laid out our general strategy, starting from this section, we will present various examples to demonstrate 
the method. 
In this section, we will consider the simplest two-loop example, i.e., the sunset topology. The integrals can be written as
\bea
I_3^{(r_1,r_2)}(n_1,n_2,n_3)&\equiv &\int d^{D}\ell_1d^{D}\ell_2{ \ell_1^{\mu_1}... \ell_1^{\mu_{r_1}}\ell_2^{\nu_1}...\ell_2^{\nu_{r_2}} \over (D_1^{n_1})(D_2^{n_2})(D_3^{n_3})}~~~~\Label{sunset-1}
\eea
with the propagators
\bea
D_1&=&\ell_1^2-m_1^2,~~D_2=\ell_2^2-m_2^2,~~D_3=(\ell_1+\ell_2-p)^2-m_3^2~~~~\Label{sunset-2}
\eea
The corresponding functions are given by
\bea
A&=&\lba{cc}
x_1+x_3&x_3\nn
x_3&x_2+x_3
\rba~~~~~
B=\lba{c}
-x_3 p\nn
-x_3 p
\rba~~~~~
C=-m_1^2x_1-m_2^2x_2+(p^2-m_3^2)x_3\nn
U(x)&=&Det[\hat A]=x_1x_2+x_1x_3+x_2x_3\nn
F(x)&=&U(x)x_4+f(x)\nn
&=&m_{1}^2 x_{1} (x_{1} (x_{2}+x_{3})+x_{2} x_{3})+m_{2}^2 x_{2} (x_{1} (x_{2}+x_{3})+x_{2} x_{3})\nn&&+m_{3}^2 x_{1} x_{2} x_{3}+m_{3}^2 x_{1} x_{3}^2+m_{3}^2 x_{2} x_{3}^2-p^2 x_{1} x_{2} x_{3}+x_{1} x_{2} x_{4}+x_{1} x_{3} x_{4}+x_{2} x_{3} x_{4}
\eea
In the following discussion, we will consider the reduction of several situations respectively, where  for simplicity we always choose the following seven scalar  integrals as the basis
\bea
I_3^{(0,0)}(2,1,1),~I_3^{(0,0)}(1,2,1),~I_3^{(0,0)}(1,1,2),~I_3^{(0,0)}(1,1,1),~I_3^{(0,0)}(1,1,0),~I_3^{(0,0)}(1,0,1),~I_3^{(0,0)}(0,1,1)~~~~~~~~\Label{massun}.
\eea

\subsection{Scalar's case}
Let us consider the reduction of scalar integrals first, i.e., $r_1=r_2=0$.
By SINGULAR, we solved the syzygy equations \eref{2.5} in lexicographical (lp) ordering directly.
There are ten generators, which are too long to write here, so we will put them  in the appendix A.
Each generator has five components, i.e., four $z_i$'s and the last one, $B$ in \eref{2.5}. One of them is
\bea
g_1&=&\Big\{0, 0, s x_3^2,
m_1^4 x_1^2 + m_2^4 x_2^2 + m_3^4 x_3^2 - m_3^2 s x_3^2 +
2 m_3^2 x_3  + x_4^2 + 2 m_2^2 x_2 (m_3^2 x_3 + x_4) \nn & & +
2 m_1^2 x_1 (m_2^2 x_2 + m_3^2 x_3 + x_4), -m_1^2 x_1 - m_2^2 x_2 -
m_3^2 x_3 - s x_3 - x_4\Big\}\label{g1}
\eea
Since the first nine are degree two and the last one, degree one, we write 
 $z_i=\frac{\sum_{j=1}^{9}c_jg_{ji}}{x_4^{2}}+\frac{c_{10}g_{10,i}}{x_4}$  as in the \eref{2.11}, and the obtained IBP relations without dimensional shift are
\bea
&&\Big\{\sum_{i=1}^{3}c_{i^{++}}i^{++}+\sum_{i,j=1;i\neq j}^{3}c_{i^{+}j^{+}}i^{+}j^{+}+\sum_{i=1}^{3}c_{i^{+}}i^{+}+\sum_{i,j=1;i\neq j}^{3}c_{i^{++}j^{-}}i^{++}j^{-}\nn
&&+\sum_{i,j=1;i\neq j}^{3}c_{i^{+}j^{-}}i^{+}j^{-}+\sum_{i,j,k=1;i\neq j\neq k}^{3}c_{i^{+}j^{+}k^{-}}i^{+}j^{+}k^{-}+c_{0,0,0}\Big\}i[\lambda_0;n_1,n_2,n_3]
+\delta_{bound}=0~~~\Label{3.6}
\eea
We can rewrite \eref{3.6} as 
\bea \sum C^{i_1,i_2, i_3}_{n_1,n_2,n_3}   i[n_1+i_1,n_2+i_2,n_3+i_3] +\delta_{bound}=0~~~~~~~~\Label{Ci1}
\eea
where the coefficients $C$'s are defined by, for example,
\bea c_{1^{++}} 1^{++} i[\lambda_0;n_1,n_2,n_3]=c_{1^{++}} i[\lambda_0;n_1+2,n_2,n_3]
\equiv C^{2,0,0}_{(n_1,n_2,n_3)}i[\lambda_0;n_1+2,n_2,n_3]\eea
Using the explicit expression of $g_i$ we can find\footnote{One can see that the $g_{10}$ is the trivial solution \eref{eu}
	and the corresponding coefficient $c_{10}$ will not appear in \eref{Ci1}. Thus it supports our claim that trivial
	solution will not contribute to nontrivial IBP relation.}
\bea
C^{2,0,0}_{n_1,n_2,n_3}&=&\frac{1}{2}m_1^2 (m_1^4 (c_2 - c_6) + s (-5 m_2^2 + 3 m_3^2 + s) c_6 \nn&&+
m_1^2 (c_1 - s c_2 + 5 m_2^2 c_6 - 3 m_3^2 c_6)) (3 D -
2 (5 +n_1 + n_2 + n_3))
\eea
here $c_1$ to $c_9$ are free parameters.
Since the coefficients are too long to write here, we put them in the appendix.

Now we discuss  the choice of free parameters $c_i$. From the expression of \eref{3.6}, we see that the allowed values  of $\{i_1,i_2, i_3\}$
in \eref{Ci1}
are given by the union of following six types of values
%
%
%
\bea
&&A_1=\cup_{Permutation}\{2,0,0\},~~A_2=\cup_{Permutation}\{1,1,0\},~~A_3=\cup_{Permutation}\{1,0,0\}\cup\{0,0,0\},\nn
&&A_4=\cup_{Permutation}\{2,-1,0\},~~A_5=\cup _{Permutation}\{1,-1,0\},~~A_6=\cup_{Permutation}\{1,1,-1\}
\eea
%
Among these six groups, we see that $A_1$ and $A_2$ shift the total power of propagators by two.
For the $A_4$ term, although the shift of total power is just one, but the power of one propagator has been shifted by two.
We could choose $x_i$ to simplify the IBP relation \eref{Ci1} by removing as many as possible groups with
shifting total power.  
For example, there is a solution of $c_i$'s such that 
\bea
\forall \{i_1,i_2,i_3\}\in A_1\cup A_4\cup A_5\cup\{\{1,0,1\},\{0,1,1\}\} ,~~ C_{n_1,n_2,n_3}^{i_1,i_2,i_3}&=&0~~~\Label{par110}
\eea
Taking the solution  in \eref{par110}, we got the simplified IBP relation
\bea
c_{1,1,0}~i[\lambda_0,n_1+1,n_2+1,n_3]+\sum_{\{i_1,i_2,i_3\}\in A_3 \cup A_6}c_{\{i_1,i_2,i_3\}} i[\lambda_0,n_1+i_1,n_2+i_2,n_3+i_3]+\delta_{1,1,0}&=&0~~~~~~\Label{ibpaft}
\eea
Notice that in the relation \eref{ibpaft}, only the first term shifts  total power by two, and all others  shift the total power at most by one. By this relation, we could  reduce $i[\lambda_0,i_1+1,i_2+1,i_3]$ to simpler integrals. Iterating the
procedure we will get the complete reduction result.  Similarly, by  different choices of the free parameters $c_1$ to $c_9$, we could get
\bea
c_{2,0,0}i[\lambda_0,n_1+2,n_2,n_3]+\sum_{\{i_1,i_2,i_3\}\in A_3 \cup A_6}c_{i_1,i_2,i_3} i[\lambda_0,n_1+i_1,n_2+i_2,n_3+i_3]+\delta_{2,0,0}&=&0\label{3.13-a}\\
c_{0,2,0}i[\lambda_0,n_1,n_2+2,n_3]+\sum_{\{i_1,i_2,i_3\}\in A_3  \cup A_6}c_{i_1,i_2,i_3} i[\lambda_0,n_1+i_1,n_2+i_2,n_3+i_3]+\delta_{0,2,0}&=&0\label{3.13-b}\\
c_{0,0,2}i[\lambda_0,n_1,n_2,n_3+2]+\sum_{\{i_1,i_2,i_3\}\in A_3  \cup A_6}c_{i_1,i_2,i_3} i[\lambda_0,n_1+i_1,n_2+i_2,n_3+i_3]+\delta_{0,0,2}&=&0\label{3.13-c}\\
c_{1,0,1}i[\lambda_0,n_1+1,n_2,n_3+1]+\sum_{\{i_1,i_2,i_3\}\in A_3  \cup A_6}c_{i_1,i_2,i_3} i[\lambda_0,n_1+i_1,n_2+i_2,n_3+i_3]+\delta_{1,0,1}&=&0\label{3.13-d}\\
c_{0,1,1}i[\lambda_0,n_1,n_2+1,n_3+1]+\sum_{\{i_1,i_2,i_3\}\in A_3  \cup A_6}c_{i_1,i_2,i_3} i[\lambda_0,n_1+i_1,n_2+i_2,n_3+i_3]+\delta_{0,1,1}&=&0~~~\Label{ibpaft2}
\eea
%
Using above  six independent IBP relations, one can quickly  reduce the power of propagators to the basis \eref{massun}\footnote{Notice that the boundary term  is different from each other. Since we chose different parameters $c_1$ to $c_9$ in each relation, the $z_1$ to $z_8$ are also different.}. The boundary terms only contribute to the sub-topology, which is the diagram removed by a inner line. We gave the analytic coefficients in the attached files. 
%
%
%
%
\subsubsection{The example: $I_3^{(0,0)}(3,1,1)$}
To get the results of $I_3^{(0,0)}(3,1,1)$ we need to reduce $i[\lambda_0;2,0,0]$. We could set $n_1=0$, $n_2=0$ and $n_3=0$ in the \eref{3.13-a}, and we will get a relation \footnote{When $n_1=n_2=n_3=0$,  the coefficients $c_{i_1,i_2,i_3}$ with $\{i_1,i_2,i_3\}$ in $A_4$ and $A_5$ are zero automatically.}
\bea
i[\lambda_0;2,0,0]&=&c_{\{0,0,0\}}i[\lambda_0;0,0,0]+c_{\{1,0,0\}}i[\lambda_0;1,0,0]+c_{\{0,1,0\}}i[\lambda_0;0,1,0]+c_{\{0,0,1\}}i[\lambda_0;0,0,1]+\delta_{boundterm}~~~~~~~~~~~\label{3.18}
\eea
which is the wanted reduction result already. To translate this result to the familiar master integrals, we need to use the correspondence between 
 $i[\lambda_0,n_1,n_2,n_3]$ and  $I_3(n_1+1,n_2+1,n_3+1)$ as given \eref{defI}, thus we get
\bea
I_3^{(0,0)}(3,1,1)&=&c_{311\to211}I_3^{(0,0)}(2,1,1)+c_{311\to 121}I_3^{(0,0)}(1,2,1)+c_{311\to112}I_3^{(0,0)}(1,1,2)\nn & & +c_{311\to111}I_3^{(0,0)}(1,1,1)+\cdots~~~~~
\eea
with the coefficients given in the Appendix A. The boundary part is just the product of two one-loop tadpoles and the corresponding 
reductions have been discussed in [cite previous work], thus we can write down the expressions directly.  
%
 The coefficients  $c_{311\to211}$, $c_{311\to121}$, $c_{311\to112}$, $c_{311\to111}$ are confirmed with the FIRE6.
 One nice point of our method is that to give the reduction results of the top-sector, we only need to use the simplified IBP relation \eref{ibpaft2} once.
 
%
\subsubsection{The example: $I_3^{0,0}(2,2,1)$}
In this case, we could set $n_1=n_2=n_3=0$ in \eref{ibpaft}, and we have the solution
\bea
i[\lambda_0,1,1,0]&=&c_{\{1,0,0\}}i[\lambda_0,1,0,0]+c_{\{0,1,0\}}i[\lambda_0,0,1,0]+c_{\{0,0,1\}}i[\lambda_0,0,0,1]+c_{\{0,0,0\}}i[\lambda_0,0,0,0]+\cdots ~~~~~
\eea
Compared to the definition of $I_3$ in \eref{defI}, we have
\bea
I_3(2,2,1)&=&c_{221\to211}I_3(2,1,1)+c_{221\to121}I_3(1,2,1)+c_{221\to112}I_3(1,1,2)+c_{221\to111}I_3(1,1,1)+\cdots `~~~~
\eea
Again, to successed reduce the top-sector to the master integrals, we need to use \eref{ibpaft} only once.
The coefficients which given in the appendix are confirmed with FIRE6.
\subsubsection{The scalar's reduction with general power}
For the general powers, we could use the recurrence relation \eref{ibpaft} to \eref{ibpaft2} to lower the total power until we reduce the top-sector  to the master integrals and get the analytic coefficients. 
For the boundary part, again we can use discussions in previous work to solve them [cite one-loop work]. 
Depending on different situations, we should choose different recurrence relations to simplify reduction procedure: 
%
\begin{itemize}
\item When One propagator has the   larger   power   than the others, for example $n_1>n_2, n_3$, we should use \eref{3.13-a}
to lower $n_1$ to $(n_1-2)$ (similarly using \eref{3.13-b} or \eref{3.13-c} to lower $n_2, n_3$).  

%
%
\item When at least two propagator have the   larger   power, for example $n_1=n_2\geq n_3$, we should use 
\eref{ibpaft} to lower $n_1, n_2$ to $(n_1-1, n_2-1)$ (and similarly using \eref{3.13-d} and \eref{ibpaft2})

%
%
\end{itemize}
According to above procedure, iteratively using \eref{ibpaft} to \eref{ibpaft2}, we will eventually get all the reduction coefficients to the master integrals. 

\subsection{Tensor's case}
%
%
Now we consider the  massive sunset with nontrivial tensor structures
\bea
I_3^{(r_1,r_2)}(n_1,n_2,n_3)&=&\int d^{D}l_1d^Dl_2\frac{l_1^{\mu_1}\cdots l_1^{\mu_{r_1}}l_2^{\nu_1}\cdots l_2^{\nu_{r_2}}}{D_1^{n_1+1}D_2^{n_2+1}D_3^{n_3+1}}
\eea
The function $F(q,y)$   is given by
\bea
F(q,y)&=&-x_1x_2x_3p^2+(p\cdot q_2x_1+p\cdot q_1x_2)x_3y+m_{1}^2 x_{1} (x_{1} (x_{2}+x_{3})\nn
&&+x_{2} x_{3})+\frac{1}{4} \Big(4 m_{2}^2 x_{2} (x_{1} (x_{2}+x_{3})+x_{2} x_{3})+4 m_{3}^2 x_{3} (x_{1} (x_{2}+x_{3})+x_{2} x_{3})\nn
&&+q_{1}^2 x_{2} y^2+q_{1}^2 x_{3} y^2-2 q_{1} q_{2} x_{3} y^2+q_{2}^2 x_{1} y^2+q_{2}^2 x_{3} y^2+4 x_{1} x_{2} x_{4}+4 x_{1} x_{3} x_{4}+4 x_{2} x_{3} x_{4}\Big)
\eea
For simplicity, we choose the scalar basis same as \eref{massun}.
\subsubsection{The example: $I_3^{(1,0)}(1,1,1)$}
%
For this case, using \eref{sumI}  we find 
\bea
i[\lambda_0,0,0,0]^{\mu_1}&=&
%
%
\frac{-\lambda_0 p^{\mu_1}}{-3+\frac{3D}{2}}i[\lambda_0-1;0,1,1]~~~\Label{322}
\eea
where the $i[\lambda_0-1,0,1,1]$ contributes to the sunset in dimension $D+2$.
To get the result of reduction in the dimension $D$, we need to reduce the $i[\lambda_0-1,0,1,1]$ to the $i[\lambda_0,\cdots]$. Different from the proposal given in [cite chen], we use the method discussed in \eref{tensortrick1},
i.e., imposing 
%
\bea
\sum_{i=1}^{n+1}g_{i}\frac{\d F}{\d x_i}&=&g_{0}x_2x_3x_4^{\alpha}~~~\Label{234a}
\eea
One point worth to explain is that to produce $i[\lambda_0-1;0,1,1]$ at the right hand side of \eref{322}, we must 
require the right hand side of \eref{234a} to be the form $g x_2 x_3$ where $g$ is only the function of $x_4$. 
By SINGULAR, we could find the solution with the choice of $\a=1$ and $g_0$ independent of $x_i$. There are 
 totally ten generators of the syzygy in lexicographical (lp) ordering and  one of them is
\bea
g_{1}&=&(4s-2m_1^2-2m_2^2-2m_3^2)x_1+(-3m_2^2)x_2-3m_3^2x_3,~~~~
g_2=(-2s-2m_1^2+m_2^2+4m_3^2)x_2+3m_3^2x_3~~~~~~~~~~~~\nn
g_3&=&3m_2^2x_2+(-2s-2m_1^2+4m_2^2+m_3^2)x_3\nn
g_4&=&(-6sm_1^2+6m_1^4)x_1+(3sm_2^2+9m_1^2m_2^2-3m_2^4-9m_2^2m_3^2)x_2\nn&&+(3sm_3^2+9m_1^2m_3^2-9m_2^2m_3^2-3m_3^4)x_3+(-2s+4m_1^2-2m_2^2-2m_3^2)x_4,~~~~g_0=-6s~~~\Label{tensorsunsetge}
\eea
 Since each of the generators from $g_1$ to $g_4$ has degree one, we could set $z_j=\frac{g_j}{x_4}, j=1,\cdots,4$ and $z_0={g_0\over x_4}$. Taking the $z_j$ into \eref{syz2} with $\la_1=\la_2=\la_3=0$, we got the recurrence relations between dimension $D+2$ to dimension $D$ as 
%
%
%
%
%
%
%
\bea
&&3 \Big(-2 (D-3) m_{1}^2+(D-3) m_{2}^2+D m_{3}^2+D s-3 m_{3}^2-2 s\Big)i[\lambda_0,0,0,0]\nn&&+\frac{3}{2} (3 D-8) m_{3}^2 \Big(-3 m_{1}^2+3 m_{2}^2+m_{3}^2-s\Big) i[\lambda_0,0,0,1]+\frac{3}{2} (3 D-8) m_{2}^2 \Big(-3 m_{1}^2+m_{2}^2+3 m_{3}^2-s\Big)i[\lambda_0,0,1,0]\nn&&-3 (3 D-8) m_{1}^2 \Big(m_{1}^2-s\Big)i[\lambda_0,1,0,0]-3Dsi[\lambda_0-1,0,1,1]+{\rm bound}=0
\eea
with the boundary term
\bea
{\rm bound}&=&3 \Big(m_{2}^2 (i[\lambda_0,-1,1,0]-i[\lambda_0,0,1,-1]+m_{3}^2 (i[\lambda_0,-1,0,1]-i[\lambda_0,0,-1,1])\Big)
\eea
Solving the $i[\lambda_0-1;0,1,1]$, we got
\bea
&&i[\lambda_0-1,0,1,1]\nn
&=&\frac{1}{D s}(-2 (D-3) m_{1}^2+(D-3) m_{2}^2+D m_{3}^2+D s-3 m_{3}^2-2 s) i[\lambda_0,0,0,0]\nn
&&+\frac{(3 D-8) m_{3}^2 \Big(-3 m_{1}^2+3 m_{2}^2+m_{3}^2-s\Big)}{2 D s} i[\lambda_0,0,0,1]
+\frac{(3 D-8) m_{2}^2 \Big(-3 m_{1}^2+m_{2}^2+3 m_{3}^2-s\Big)}{2 D s} i[\lambda_0,0,1,0]\nn
&&+-\frac{(3 D-8) m_{1}^2 \Big(m_{1}^2-s\Big)}{D s} i[\lambda_0,1,0,0]+\frac{m_{3}^2}{D s} i[\lambda_0,-1,0,1]\nn&&-\frac{m_{3}^2}{D s}i[\lambda_0,0,-1,1]
+\frac{m_{2}^2}{D s}i[\lambda_0,-1,1,0]-\frac{m_{2}^2}{D s} i[\lambda_0,0,1,-1]~~~\Label{436}
\eea
Here the $i[\lambda_0,-1,0,1]$, $i[\lambda_0,0,1,-1]$ and $i[\lambda_0,0,-1,1]]$  come from the boundary terms. Notice that the right hand of the \eref{436} all contribute to the basis we chose, which means we got the results of reduction by solving syzygy equation once. Combining \eref{436} and \eref{322} with \eref{defI}, we got the results of reduction
\bea
I_3^{(1,0)}(1,1,1)&=&c_{10\to211}I_3^{(0,0)}(2,1,1)+c_{10\to121}I_3^{(1,0)}(1,2,1)+c_{10\to112}I_3^{(1,0)}(1,1,2)\nn&&+c_{10\to111}I_3^{(1,0)}(1,1,1)+\cdots ~~~~~~~~~~
\eea
with the coefficients (without the boundary terms)
\bea
c_{10\to211}&=&\frac{(-8+3D)m_1^2(s-m_1^2)p^{\mu_1}}{3(D-2)s}\nn
c_{10\to121}&=&\frac{(-8+3D)m_2^2(-3m_1^2+m_2^2+3m_3^2-s)p^{\mu_1}}{6(D-2)s}\nn
c_{10\to112}&=&\frac{(-8+3D)m_3^2(-3m_1^2+3m_2^2+m_3^2-s)p^{\mu_1}}{6(D-2)s}\nn
c_{10\to111}&=&\frac{(3D-8)\Big(-2(D-3)m_1^2+(D-3)m_2^2+(D-3)m_3^2+(D-2)s\Big)}{6(D-2)s}p^{\mu_1}
\eea
The coefficients are confirmed with FIRE6 \cite{Smirnov:2019qkx}. Here we just got the reduction result of the top-sector. To get the complete results, we need to reduce the boundary terms by repeating the similar procedure above.
\subsubsection{The example: $I_3^{(1,1)}(1,1,1)$}
Now let us consider a more complex case, i.e., tensor rank $(1,1)$. By directly calculation, we got
\bea
i[\lambda_0,0,0,0]^{\mu_1\nu_1}&=&
\frac{\lambda_0}{\Gamma(3)}\frac{g^{\mu_1\nu_1}}{(\lambda_4+1)(\lambda_4+2)}\Big\{i[\lambda_0-1;0,1,0]+i[\lambda_0-1;0,0,1]\Big\}\nn
&&+\frac{\lambda_0(\lambda_0-1)}{\Gamma(3)}\frac{2p^{\mu_1}p^{\nu_1}}{(\lambda_4+1)(\lambda_4+2)}i[\lambda_0-2;0,2,2]~~\Label{iuv}
\eea
The two terms in the first line of \eref{iuv} contribute to the integrals in dimension $D+2$, the last term in the second line contributes to the integral in dimension $D+4$. To reduce $i[\lambda_0-2;0,2,2]$, we could solve
$
\sum_{j=1}^4 g_{j}\frac{\d F}{\d x_j}=g_0 x_2^2x_3^2x_4^{\a}
$. 
Choosing $\a=0$, we got one of the generators
\bea
g_1&=&2 x_{3} (m_3^2 x_{3}+x_{4})
(m_1^2 (2 m_2^2-3 s)+m_1^4-2
(m_2^2-s)
(m_2^2-m_3^2+s))\nn
g_2&=&2 m_1^2 (-m_2^2 (m_3^2 (2
x_{3} (x_{3}-x_{1})+3 x_{2}^2+2 x_{2}
x_{3})+s (2 x_{1} (x_{2}+x_{3})-x_{2}
(9 x_{2}+2 x_{3}))+2 x_{4} (2
x_{2}+x_{3}))\nn&&+m_2^4 (x_{1}+x_{2})
(x_{2}-x_{3})+s (m_3^2 (-2 x_{1}
x_{3}+3 x_{2}^2+4 x_{2} x_{3}+3
x_{3}^2)-s (x_{1} (3
x_{2}+x_{3})\nn&&+x_{2} (12 x_{2}+7 x_{3}))+3
x_{4} (2 x_{2}+x_{3})))+m_1^4
(m_2^2 (-x_{1} x_{2}+3 x_{1}
x_{3}-8 x_{2}^2+3 x_{2} x_{3})-2 (2
x_{2}+x_{3}) (m_3^2
x_{3}+x_{4})\nn&&+s (15 x_{1} x_{2}+9
x_{1} x_{3}+15 x_{2}^2+13 x_{2}
x_{3}))+m_1^6 (-(6 x_{1}
x_{2}+4 x_{1} x_{3}+3 x_{2}^2+4 x_{2}
x_{3}))\nn&&+4 (m_2^2-s)
(m_2^2-m_3^2+s) (m_2^2
x_{2}^2+(2 x_{2}+x_{3}) (m_3^2
x_{3}+x_{4})-s x_{2} (3 x_{2}+2
x_{3}))\nn
g_3&=&m_1^4 (m_2^2 (x_{1} (x_{2}-3
x_{3})+x_{3} (x_{2}+4 x_{3}))-2 x_{3}
(m_3^2 x_{3}+x_{4})-s (15
x_{1} x_{2}+9 x_{1} x_{3}+15 x_{2}
x_{3}+x_{3}^2))\nn&&+2 m_1^2
(m_2^2 (m_3^2 (-x_{3}) (2
x_{1}+x_{3})+2 s x_{1} (x_{2}+x_{3})+s
x_{3} (2 x_{2}-x_{3})-2 x_{3}
x_{4})\nn&&+m_2^4 (x_{1}+x_{3})
(-(x_{2}-x_{3}))+s (2 m_3^2 x_{3}
(x_{1}+x_{3})+s x_{1} (3 x_{2}+x_{3})+s
x_{3} (3 x_{2}-2 x_{3})+3 x_{3}
x_{4}))\nn&&+m_1^6 (6 x_{1}
x_{2}+4 x_{1} x_{3}+6 x_{2}
x_{3}+x_{3}^2)-4 x_{3}
(m_2^2-s)
(m_2^2-m_3^2+s) (m_2^2
x_{3}-m_3^2 x_{3}+s x_{3}-x_{4})\nonumber
\eea
\bea
g_4&=&-2 m_1^2 (-m_2^2 (m_3^2
(x_{4} (4 x_{1}+3 x_{2}+5 x_{3})-s
(10 x_{1} x_{2}-2 x_{1} x_{3}+6
x_{2}^2+23 x_{2} x_{3}+x_{3}^2))\nn&&+3
m_3^4 x_{3} (2 x_{1}+x_{2})+s^2 (7
x_{1} x_{2}-3 x_{1} x_{3}+21 x_{2}^2+13
x_{2} x_{3})-s x_{4} (23 x_{2}\nn&&+2
x_{3})+4 x_{4}^2)-m_2^4 (m_3^2
(3 x_{1} (x_{2}-3 x_{3})+6 x_{2}^2+4
x_{2} x_{3}+3 x_{3}^2)+s (7 x_{1}
x_{2}+3 x_{1} x_{3}-20 x_{2}^2\nn&&+6 x_{2}
x_{3})+4 x_{4} (-x_{1}+2
x_{2}+x_{3}))+m_2^6 (3 x_{1}+x_{2})
(x_{2}-x_{3})+s (m_3^2 (x_{4} (4
x_{1}+3 x_{2}+10 x_{3})-3 s (x_{1}
(x_{2}+x_{3})\nn&&+x_{3} (3 x_{2}+2 x_{3})))+3
m_3^4 x_{3} (2 x_{1}+x_{2}+x_{3})+3 s^2
(x_{1} (x_{2}+x_{3})+4 x_{2} x_{3})-s
x_{4} (4 x_{1}+9 x_{2}+4 x_{3})+6
x_{4}^2))\nn&&+m_1^4 (m_2^2
(m_3^2 (5 x_{1} (x_{2}+x_{3})+x_{3}
(14 x_{2}-13 x_{3}))-s x_{2} (36 x_{1}+15
x_{2}+25 x_{3})+x_{4} (8 x_{1}+15 x_{2}-9
x_{3}))\nn&&+m_2^4 (x_{1} (x_{2}-3
x_{3})+5 x_{2} (3 x_{2}-2 x_{3}))+m_3^2
(3 s (x_{1} (3 x_{2}+x_{3})+x_{3} (5
x_{2}-x_{3}))\nn&&+10 x_{3} x_{4})+6 m_3^4
x_{3}^2+3 s^2 x_{1} x_{2}+15 s^2 x_{1}
x_{3}+15 s^2 x_{2} x_{3}-12 s x_{1} x_{4}-4
s x_{3} x_{4}+4 x_{4}^2)\nn&&+m_1^6
(m_2^2 (13 x_{1} x_{2}-3 x_{1}
x_{3}+7 x_{2} x_{3})-3 m_3^2 x_{2} (2
x_{1}+3 x_{3})+6 s x_{1} x_{2}-12 s x_{1}
x_{3}-3 s x_{2} x_{3}\nn&&+4 x_{1} x_{4}-3
x_{2} x_{4}+x_{3} x_{4})+3 m_1^8
x_{1} (x_{3}-x_{2})-4 (m_2^2-s)
(m_2^2-m_3^2+s) (-m_2^2
(m_3^2 x_{3} (x_{3}\nn&&-5 x_{2})+2 s
x_{2} (3 x_{2}+x_{3})-5 x_{2}
x_{4})+m_2^4 x_{2} (2
x_{2}-x_{3})+(m_3^2 x_{3}-s
x_{3}+x_{4}) (3 m_3^2 x_{3}-3 s
x_{2}+2 x_{4}))\nonumber
\eea
\bea
g_0&=&3 s (m_1^2-2 (m_2^2+s)) (m_1^2 (2 m_2^2-3 s)+m_1^4-2 (m_2^2-s) (m_2^2-m_3^2+s))~~~~\Label{syzoftensor11}
\eea
Taking the \eref{syzoftensor11} into \eref{syz2}, we reduce the $i[\lambda_0-2;0,2,2]$ to integral in dimension $\lambda_0-1=-\frac{D+2}{2}$. Notice the $\lambda_0$ in \eref{syz2} should be replaced by $\lambda_0-1$. We have the result
\bea
&&i[\lambda_0-2;0,2,2]\nn
&=&c_{i;022\to000}i[\lambda_0-1;0,0,0]+c_{i;022\to100}i[\lambda_0-1;1,0,0]+c_{i;022\to010}i[\lambda_0-1;0,1,0]+c_{i;022\to001}i[\lambda_0-1;0,0,1]\nn
&&+c_{i;022\to110}i[\lambda_0-1;1,1,0]+c_{i;022\to101}i[\lambda_0-1;1,0,1]+c_{i;022\to011}i[\lambda_0-1;0,1,1]+c_{i;022\to200}i[\lambda_0-1;2,0,0]\nn
&&+c_{i;022\to020}i[\lambda_0-1;0,2,0]+c_{i;022\to002}i[\lambda_0-1;0,0,2]+\cdots~~~~~~~~\Label{334}
\eea
with the coefficients given in appendix.
Now we have reduced the situation to the similar one studies in previous subsubsection (see \eref{322}). 
Using the similar method we can finish the reduction completely.


%
\subsubsection{The example: $I_3^{(1,0)}(2,1,1)$}
This example has one tensor structure  as well as a high power propagator. Again we have 
\bea
i[\lam_0;1,0,0]^{\mu_1} 
%
%
&=&\frac{D}{3D-8}p^{\mu_1}i[\lam_0-1;1,1,1]~~~\Label{imu1100}
\eea
To reduce the $i[\lam_0,1;1,1,1]$, we could  use the solution of the syzygy equation  \eref{tensorsunsetge} in the first example. Choosing the same $z_j=\frac{g_j}{x_4}$ and $\lam_1=1$, $\lam_2=\lam_3=0$ in the recurrence relation \eref{syz2}, 
%
%
%
%
After some expansion 
we could get the recurrence relation
\bea
&&I_3^{(1,0)}(2,1,1)\nn
&=&c_{t;211\to211}I_3^{(0,0)}(2,1,1)+c_{t;211\to212}I_3^{(0,0)}(2,1,2)+c_{t;211\to221}I_3^{(0,0)}(2,2,1)\nn
&&+c_{t;211\to112}I_3^{(0,0)}(1,1,2)+c_{t;1,2,1}I_3^{(0,0)}(1,2,1)+c_{t;211\to311}I_3^{(0,0)}(3,1,1)+\cdots
\eea
with the coefficients
\bea
c_{t;211\to211}&=&\frac{\Big((4-D)m_1^2+(D-3)m_2^2+(D-3)m_3^2+(D-4)s\Big)p^{\mu_1}}{(3D-8)s}\nn
c_{t;211\to212}&=&\frac{m_3^2(-3m_1^2+3m_2^2+m_3^2-s)p^{\mu_1}}{(3D-8)s}\nn
c_{t;211\to221}&=&\frac{m_2^2(-3m_1^2+m_2^2+3m_3^2-s)p^{\mu_1}}{(3D-8)s}\nn
c_{t;211\to112}&=&\frac{m_3^2p^{\mu_1}}{(3D-8)s},~~
c_{t;211\to121}=\frac{m_2^2p^{\mu_1}}{(3D-8)s}\nn
c_{t;211\to311}&=&\frac{4m_1^2(s-m_1^2)p^{\mu_1}}{(3D-8)s}~~~\Label{I310211}
\eea
Notice that not all of the six basis in the right hand side of \eref{I310211} are the master integrals, and the three terms $I_3^{(0,0)}(2,1,2)$, $I_3^{(0,0)}(2,2,1)$, $I_3^{(0,0)}(3,1,1)$ are still needed to be reduced. After reducing the three terms and adding the boundary terms, we got the final results which confirmed with FIRE6.
%
%

Having above examples, we can see that for general integrals with arbitrary tensor structures and arbitrary powers,
first we need to use the differential operators \eref{tensor2} to get the expression \eref{sumI}, where each one is
the scalar integrals, but possibly with different dimension. Secondly, we need to shift dimension back by the trick
\eref{tensortrick1}. After these two steps, we are left with only scalar integrals in the same dimension, but 
different powers. Now we can use the trick \eref{syzy} to reduce these scalar integrals to the master basis.


%
%
\section{Example 2: double triangle diagram}
Now we consider an more complicated example, 
\bea
I_5^{(r_1,r_2)}(n_1,\cdots n_5)&=&\int d^{D}l_1d^{D}l_2  \frac{l_1^{\mu_1}\cdots l_1^{\mu_{r_1}}l_2^{\nu_2}\cdots l_1^{\nu_{r_2}}}{D_1^{n_1}\cdots D_5^{n_5}}
\eea
with the propagators 
\bea
D_1&=&l_1^2-m_1^2,~~D_2=l_2^2-m_2^2,~~D_3=(l_1+l_2)^2-m_3^2,~~D_4=(l_1-p)^2-m_4^2,~~D_5=(l_2+p)^2-m_5^2~~~~~~~~~~~~~~~
\eea
The corresponding functions are
\bea
F&=&x_1 \Big(x_2 \Big(x_3
(m_{1}^2+m_{2}^2+m_{3}^2)+x_4
(m_{1}^2+m_{4}^2-s)+m_{2}^2 x_5+m_{5}^2
x_5-s x_5+x_6\Big)\nn&&+x_3 \Big(x_5
(m_{1}^2+m_{3}^2+m_{5}^2-s)+x_4
(m_{1}^2+m_{4}^2-s)+x_6\Big)+x_5 (x_4
(m_{1}^2+m_{4}^2-s)\nn&&+m_{5}^2
x_5+x_6)+m_{2}^2 x_2^2+m_{3}^2 x_3^2\Big)+m_{1}^2
(x_2+x_3+x_5) x_1^2+x_2 \Big(x_3 \Big(x_4
(m_{2}^2+m_{3}^2\nn&&+m_{4}^2-s)+x_5
(m_{2}^2+m_{5}^2-s)+x_6\Big)+x_4 \Big(x_5
(m_{2}^2+m_{5}^2-s)+m_{4}^2
x_4+x_6\Big)+m_{3}^2 x_3^2\Big)\nn&&+m_{2}^2 x_2^2
(x_3+x_4)+\Big(x_4 x_5+x_3 (x_4+x_5)\Big)
\Big(m_{3}^2 x_3+m_{4}^2 x_4+m_{5}^2 x_5+x_6\Big)\nn
A&=&\left[\begin{array}{cc}
	x_1+x_3+x_4&x_3\nn
	x_3&x_2+x_3+x_5
	\end{array}\right]
,~~B=\left[\begin{array}{c}
	-px_4\nn
	px_5
	\end{array}\right]\nn
C&=&-m_1^2x_1-m_2^2x_2-m_3^2x_3-m_4^2x_4-m_5^2z_5+p^2(x_4+x_5)~~~\Label{4.3}
\eea
For simplicity, we always choose the scalar basis $I_{5}^{(0,0)}(1,1,1,1,1)$ as the basis of top-sector in the following calculation.
\bea
\eea
\subsection{Scalar's case}
In the scalar's case, i.e., $r_1=r_2=0$, we solved the syzygy equations \eref{syzy} using degrevlex ordering, and got 8 generators of the syzygy with degree one. Using the similar method in sunset's case, we define $z_i\equiv \frac{\sum_{i=1}^{8}c_ig_{ji}}{x_6}$ and obtain the following IBP recurrence relation without dimensional shift,
\bea
\sum_{\{i_1,\cdots,i_5\}\in A_0\cup\cdots \cup A_3} C_{n_1,\cdots, n_5}^{i_1\cdots i_5} i[n_1+i_1,\cdots n_5+i_5]+\delta_{bound}&=&0
\eea
where  the four types of the values of $\{i_1,\cdots ,i_5\}$ are
\bea
A_0&=&\{0,0,0,0,0\},~~A_1=\cup_{Permutation}\{1,0,0,0,0\},\nn
A_2&=&\cup_{Permutation}\{1,-1,0,0,0\},~~A_3=\cup_{Permutation}\{-1,0,0,0,0\}~~~\Label{doubletriangle}
\eea
By choosing the particular values of $c_1$ to $c_8$, we could get five independent recurrence relation to lower the total power,
\bea
&&c_{1,0,0,0,0} i[\lam_0,n_1+1,\cdots,\cdots n_5]+\sum_{\substack{\{k_1,\cdots,k_5\}\\ \in A_0\cup A_2\cup A_3}}c_{k_1,\cdots,k_5} i[\lam_0,n_1+k_1,\cdots n_5+k_5]+\delta_{B,1}=0~~\Label{line1dbt}\\
&&c_{0,1,0,0,0} i[\lam_0,n_1,\cdots,n_2+1,\cdots n_5]+\sum_{\substack{\{k_1,\cdots,k_5\}\\ \in A_0\cup A_2\cup A_3}}c_{k_1,\cdots,k_5} i[\lam_0,n_1+k_1,\cdots n_5+k_5]+\delta_{B,2}=0~~\Label{line2dbt}\\
&&c_{0,0,1,0,0} i[\lam_0,n_1,\cdots,n_3+1,\cdots n_5]+\sum_{\substack{\{k_1,\cdots,k_5\}\\ \in A_0\cup A_2\cup A_3}}c_{k_1,\cdots,k_5} i[\lam_0,n_1+k_1,\cdots n_5+k_5]+\delta_{B,3}=0~~\Label{line3dbt}\\
&&c_{0,0,0,1,0} i[\lam_0,n_1,\cdots,n_4+1,\cdots n_5]+\sum_{\substack{\{k_1,\cdots,k_5\}\\ \in A_0\cup A_2\cup A_3}}c_{k_1,\cdots,k_5} i[\lam_0,n_1+k_1,\cdots n_5+k_5]+\delta_{B,4}=0~~\Label{line4dbt}\\
&&c_{0,0,0,0,1} i[\lam_0,n_1,\cdots,\cdots n_5+1]+\sum_{\substack{\{k_1,\cdots,k_5\}\\ \in A_0\cup A_2\cup A_3}}c_{k_1,\cdots,k_5} i[\lam_0,n_1+k_1,\cdots n_5+k_5]+\delta_{B,5}=0~~~~~~~\Label{line5dbt}
\eea
where the boundary terms $\delta_{B,j}$ 
 are too long to be shown here.
\subsubsection{The example: $I_5^{(0,0)}(2,1,1,1,1)$}
In this example, we need to reduce $i[\lam_0;1,0,0,0,0]$. We could use the first line in \eref{line1dbt} with $n_1=\cdots=n_5=0$ and solve the target $i[\lam_0;1,0,0,0,0]$.  The result is
\bea
i[\lam_0;1,0,0,0,0]&=&c_{10000\to00000}i[\lam_0;0,0,0,0,0]+\cdots
\eea
with the coefficients
\bea
c_{10000\to00000}&=&\frac{-2m_1^2m_5^2+m_2^2(-m_3^2+m_4^2+m_5^2)+m_3^2m_5^2 +m_3^2 s +m_4^2 m_5^2 -m_4^2 s +m_5^2 s -m_5^4}{N_{11111}}\nn
N_{11111}&=&3 (m_1^2 (m_2^2(m_3^2-m_4^2-m_5^2 ) -m_3^2(m_5^2+s)  -(m_4^2-m_5^2)(m_5^2-s) ) +m_1^4m_5^2 -m_2^2 (m_3^2(m_4^2+s) \nn
&&-(m_4^2-m_5^2)(m_4^2-s)) +m_2^4m_4^2 +m_3^2 (s (m_3^2-m_5^2+s)+m_4^2(m_5^2-s))  )
\eea
Then using the \eref{defI} to translate $i[\lam_0;n_1,\cdots,n_5]$ to $I[n_1+1,\cdots,n_5+1]$, we got the reduction coefficients which is confirmed with FIRE6. Here we just used the simplified IBP relation {\bf once}.

\subsubsection{The example: $I_5^{(0,0)}(3,1,1,1,1)$}
In this case, we need to reduce $i[\lam_0;2,0,0,0,0]$. In the first step, one could reduce $i[\lam_0;2,0,0,0,0]$ to terms having {\bf total lower power} by \eref{line1dbt}, with the results,
\bea
i[\lam_0;2,0,0,0,0]&=&c_{2;10000}i[\lam_0;1,0,0,0,0]+c_{2;01000}i[\lam_0;0,1,0,0,0]+c_{2;00100}i[\lam_0;0,0,1,0,0]\nn
&&+c_{2;00010}i[\lam_0;0,0,0,1,0]+c_{2;00001}i[\lam_0;0,0,0,0,1]+c_{2;00000}i[\lam_0;0,0,0,0,0]+\cdots 
\eea
To get the final reduction results of  $i[\lam_0;2,0,0,0,0]$, we should go on and  reduce $i[\lam_0;1,0,0,0,0]$, $i[\lam_0;0,1,0,0,0]$, $i[\lam_0;0,0,1,0,0]$, $i[\lam_0;0,0,0,1,0]$ and $i[\lam_0;0,0,0,0,1]$ to $i[\lam_0;0,0,0,0,0]$. By using  \eref{line1dbt} to \eref{line5dbt}, we could easily get the reduction coefficients from these five terms to $i[\lam_0;0,0,0,0,0]$. The final result is confirmed with FIRE6.
\subsubsection{The scalar's reduction with general power}
For the case with general powers in double-triangle, things are similar. Iteratively using  the  five recurrence relations \eref{line1dbt} to \eref{line5dbt}, one can lower the total power one by one to the scalar basis as did in previous two examples. 
\subsection{Tensor's case}
%
%
For the tensor case, the function $F(q,y)$ is 
\bea
F(q,y)&=&F(0,0)+y\Big(x_2(x_3+x_4+p \cdot q_1 x_4) + x_1 (x_2+x_3 +x_5-p \cdot q_2 x_5)\nn&& +(1+p\cdot (q_1-q_2)) (x_4 x_5 +x_3(x_4+x_5)) \Big)\nn
&&-\frac{1}{4}y^2\Big( -2q_1\cdot q_2 x_3+q_2^2 (x_1+x_3+x_4)+ q_1^2 (x_2+x_3+x_5)  \Big)
\eea
where $F(0,0)$ is the function $F$   given in \eref{4.3}.

\subsubsection{The example: $I_5^{(1,0)}(1,1,1,1,1)$}
By \eref{tensor1}, we got 
\bea
i[\lam_0;0,0,0,0,0]^{\mu_1}&=&\frac{\frac{D}{2}}{(\frac{3D}{2}-6)(\frac{3D}{2}-5)} p^{\mu_1} \Big(i[\lam_0-1;0,1,0,1,0]+i[\lam_0-1;0,0,0,1,1]\nn&&+i[\lam_0-1;0,0,1,1,0] +i[\lam_0-1;0,0,1,0,1]  \Big)~~~~~\Label{tensorexample10}
\eea
To get the reduction in the  dimension $D$, we need to reduce the four temrs in dimension $D+2$ using the method in \eref{tensortrick1}.  By solving syzygy equation
\bea
\sum_{i=1}^{6} g_i\frac{\d F}{\d x_i}&=&g_0 (x_2x_4+x_4x_5+x_3x_4+x_3x_5)x_6^{\alpha}
\eea
using SINGULAR (Here we choose $\alpha=0$), we got the solutions where one generator reads
\bea
g_1&=&1,~~g_2=0,~~g_3=0,~~g_4=-1,~~g_5=0,~~g_6=-m_1^2+m_4^2-s,~~g_0=-2s
\eea
Using the $g_0$ to $g_6$ in the IBP identity \eref{syz2} with $z_i\equiv g_i$, we have
\bea
&&-2s \Big(i[\lam_0-1;0,1,0,1,0]+ i[\lam_0-1;0,0,0,1,1] +i[\lam_0-1;0,0,1,1,0] +i[\lam_0-1;0,0,1,0,1] \Big) \nn&&+\frac{1}{2}(10-3D)(m_1^2-m_4^2+s) i[\lam_0;0,0,0,0,0] +  boundary~term = 0 ~~~\Label{4.20}
\eea
Taking \eref{4.20} into \eref{tensorexample10}, we solved the $i[\lam_0;0,0,0,0,0]^{\mu_1}$ in dimension $D$. Then by \eref{sumI}, we got the reduction result
\bea
I[\lam_0;1,1,1,1,1]^{\mu_1}&=& \frac{p^{\mu_1} (m_1^2-m_4^2+s)}{2s} I_5^{(0,0)}(1,1,1,1,1)
\eea
where the coefficients are confirmed with FIRE6.
\subsubsection{The example: $I_5^{(1,1)}(1,1,1,1,1)$ (to be checked)}
By directly calculation, 
\bea
&&i[\lam_0;0,0,0,0,0]^{\mu\nu}\nn
&=&\frac{(\lam_0 g^{\mu\nu})}{2}\times\frac{2}{\lam_6(\lam_6+1)(\lam_6+2)}\int d\Pi^{(6)} F^{\lam_0-1}x_3x_6^{\lam_6+2}-\frac{\lam_0(\lam_0-1)p^{\mu}p^{\nu}}{2} \frac{2}{\lam_6(\lam_6+1)(\lam_6+2)}\nn
&&\times \int d\Pi^{(6)}F^{\lam_0-2}\Big(x_2x_4+x_4x_5+x_3(x_4+x_5)\Big)\Big( (x_1+x_4)x_5+x_3(x_4+x_5)\Big)x_6^{\lam_6+2}~~~~~~~~
\eea
where the terms in the second line contribute to Feynman integrals in dimension parameter $\lam_0-2$.
By syzygy trick  disscussed above, we should  first reduce the integrals in dimension parameter $\lam_0-2$ to dimension parameter $\lam_0-1$. To achieve this, we should solve the following syzygy equation,
\bea
\sum_{i=1}^{6}g_i\frac{\d F}{\d x_i}&=&g_0\Big(x_2x_4+x_4x_5+x_3(x_4+x_5)\Big)\Big( (x_1+x_4)x_5+x_3(x_4+x_5)\Big) x_6^{\alpha}
\eea
By choosing $\a=0$, we got the following result
\bea
g_1&=&0,~~g_2=m_4^2x_4^2-m_3^2(x_3+x_4)x_5+x_4(m_5^2x_5+x_6),g_3=x_4(m_2^2x_3-m_3^2x_3-2sx_3-sx_4+m_1^2(x_1+x_3+x_4))\nn
g_4&=&x_4\Big(-((m_1^2+2(-m_2^2+m_3^2+s))x_1-(m_1^2-m_2^2+m_3^2)x_3-(m_1^2-2m_2^2+2m_3^2+s)x_4\Big)\nn
g_5&=&-m_1^2x_1x_4-2sx_2x_4+(m_1-m_2^2+m_3^2)x_3x_4-(m_1^2+m_4^2-s)x_4^2+m_3^2x_3x_5+(m_3^2-m_5^2)x_4x_5-x_4x_6\nn
g_6&=&(2(m_2^2-m_3^2-s)(s-m_4^2)+m_1^2(-2m_2^2+m_3^2+m_4^2+m_5^2+2s))x_1x_4-2(m_2^4+s(-m_5^2+s)-m_2^2(m_3^2\nn
&&+2s))x_2x_4+(m_2^2(-3m_3^2-m_4^2+m_5^2)-m_1^2(m_3^2-m_4^2+m_5^2)+m_3^2(3m_3^2+m_4^2-m_5^2+6s))x_3x_4+(-5m_2^2m_4^2\nn
&&+4m_3^2m_4^2+m_4^2m_5^2+m_1^2(-m_3^2+m_4^2+m_5^2)+m_3^2s+4m_4^2s-m_5^2s)x_4^2+m_3^2(m_2^2-m_5^2+s)x_3x_5\nn
&&+(m_5^4+m_2^2(m_3^2-3m_5^2)+3m_5^2s+m_3^2(m_5^2+s))x_4x_5+(-3m_2^2+2m_3^2+m_5^2+3s)x_4x_6\nn
g_0&=&2m_3^2 s
\eea
Defing $z_i\equiv \frac{g_i}{x_6^2}$ in the \eref{syz2} with $\lam_1=\cdots=\lam_5=0$, and  replace $\lam_0$ by $\lam_0-1$. By similar procedure, we reduce the integrals in dimension parameter $\lam_0-1$ to dimension parameter $\lam_0$. 

\section{Example 3: Topology F in \cite{Xu:2018eos}}
This example is given by
\bea
I_7^{(r_1,r_2)}(n_1,\cdots n_7)&=&\int d^D l_1 d^D l_2\frac{l_1^{\mu_1}\cdots l_1^{\mu_{r_1}} l_1^{\mu_2}\cdots l_2^{\mu_{r_2}}  }{D_1^{n_1}\cdots D_{7}^{n_7}}
\eea
where the propagators are given by \footnote{In \cite{Needed} there are $D_8=(l_1+p_3)^2$ and $D_9=(l_2-p_1)^2-m^2$ which is for the tensor structure. Since we use the different way to deal with tensor, we do not include them here.}
\bea
\{D_1,\cdots D_7\}&=&\{(l_1-p_1)^2,l_1^2,(l_1+p_2)^2,(l_1+l_2-p_1)^2-m^2,l_2^2-m^2,\nn
&&(l_2+p_3)^2-m^2,(l_1+l_2+p_2+p_3)^2-m^2,(l_1+p_3)^2,(l_2-p_1)^2-m^2\}
\eea
All external momenta are set to be on-shell, so the kinematic variables are 
\bea
s&=&(p_1+p_2)^2,~~t=(p_1+p_3)^2,~~u=(p_2+p_3)^2=(p_1+p_4)^2=-s-t,~~p_1^2=p_2^2=p_3^2=p_4^2=0
\eea

%

%
The corresbonding functions are given by
\bea
A&=&\lba{cc}
x_1+x_2+x_3+x_4+x_7&x_4+x_7\nn
x_4+x_7&x_4+x_5+x_6+x_7\nn
\rba\nn
B&=&\lba{c}
-(x_1+x_4)p_1+(x_3+x_7)p_2+x_7p_3\nn
-x_4p_1+x_7p_2+(x_6+x_7)p_3
\rba\nn
C&=&-m^2(x_4+x_5+x_6+x_7)+p_{1}^2 (x_{1}+x_{4})+p_{2}^2 (x_{3}+x_{7})+2 p_{2}\cdot p_{3} x_{7}+p_{3}^2 x_{6}+p_{3}^2 x_{7}
\eea
The function F is given by
\bea
F(x)&=&U(x)x_8+f(x)\nn
&=&m^2 (x_{4}+x_{5}+x_{6}+x_{7}) (x_{1} (x_{4}+x_{5}+x_{6}+x_{7})+x_{2} (x_{4}+x_{5}+x_{6}+x_{7})+x_{3} x_{4}+x_{3} x_{5}\nn&&+x_{3} x_{6}+x_{3} x_{7}+x_{4} x_{5}+x_{4} x_{6}+x_{5} x_{7}+x_{6} x_{7})-s (x_{1} x_{3} (x_{4}+x_{5}+x_{6}+x_{7})+x_{1} x_{6} x_{7}\nn&&+x_{5} (x_{3} x_{4}-x_{2} x_{7}))+t x_{2} (x_{5} x_{7}-x_{4} x_{6})+x_{8} (x_{1} (x_{4}+x_{5}+x_{6}+x_{7})+x_{2} (x_{4}+x_{5}+x_{6}+x_{7})\nn&&+x_{3} x_{4}+x_{3} x_{5}+x_{3} x_{6}+x_{3} x_{7}+x_{4} x_{5}+x_{4} x_{6}+x_{5} x_{7}+x_{6} x_{7})
\eea
For later convenience, we choose the scalar basis in the top topology as
\bea
&&I_7^{(0,0)}(2,1,1,1,1,1,1),~~I_7^{(0,0)}(1,1,1,2,1,1,1),~~I_7^{(0,0)}(1,1,1,1,2,1,1),\nn
&&I_7^{(0,0)}(1,1,1,1,1,2,1),~~I_7^{(0,0)}(1,1,1,1,1,1,2),~~I_7^{(0,0)}(1,1,1,1,1,1,1),~~\cdots
\eea
where the "$\cdots$" represents the basis in the subtopologies.
\subsection{Scalar's case}
In the scalar's case, we solved the syzygy equations \eref{syzy} in lexicographical order and got 55 generators, where 2 in degree one, 37 in degree two, and 16 in degree three. Using the similar method in the former two examples, we obtain the following IBP recurrence relations without dimensional shift,
\bea
\sum_{\{i_1,\cdots ,i7\}\in A_0\cup A_{12}}C_{n_1,\cdots,n_7}^{i_1,\cdots,i_7} i[\lam_0;n_1+i_1,\cdots,n_7+i_7]+\delta_{bound}&=&0
\eea
with 13 types of the values of $\{i_1,\cdots ,i_7\}$ as
\bea
A_1&=&\cup_{Permutation}\{3,0,0,0,0,0,0\},~~A_2=\cup_{Permutation}\{2,1,0,0,0,0,0\},~~
A_3=\cup_{Permutation}\{1,1,1,0,0,0,0\}\nn
A_4&=&\cup_{Permutation}\{2,0,0,0,0,0,0\},~~
A_5=\cup_{Permutation}\{1,1,0,0,0,0,0\},~~A_6=\cup_{Permutation}\{3,-1,0,0,0,0,0\}\nn
A_7&=&\cup_{Permutation}\{2,1,-1,0,0,0,0\},~~
A_8=\cup_{Permutation}\{1,1,1,-1,0,0,0\},~~A_9=\cup_{Permutation}\{1,0,0,0,0,0,0\}\nn
A_{10}&=&\cup_{Permutation}\{2,-1,0,0,0,0,0\},~~A_{11}=\cup_{Permutation}\{1,1,-1,0,0,0,0\},~~A_{12}=\cup_{Permutation}\{1,-1,0,0,0,0,0\}\nn
A_{0}&=&\{0,0,0,0,0,0,0\}
\eea
Note there are three types of generators of syzygy in degree one, two, and three respectively. The generators in degree one read as
\bea
g_{1i}&=&\Big\{x_2+x_3,-x_2,2 x_1+2 x_2+x_3,x_7,x_6,2 x_5+x_6,2 x_4+x_7\nn&&,-2 \Big(3 m^2 x_4+3 m^2 x_5+3 m^2 x_6+3 m^2 x_7+2 x_8 \Big)+2 s x_1+s x_3\Big\}\nn
g_{2i}&=&\Big\{2 x_1+3 \Big(x_2+x_3\Big),-x_2,-x_3,2 x_4+3 x_7,2 x_5+3 x_6,-x_6,-x_7,\nn&&3 s x_3-2 \Big(3 m^2 x_4+3 m^2 x_5+3 m^2 x_6+3 m^2 x_7+2 x_8\Big)\Big\}
\eea
Defining 
\bea
z_{k}&\equiv& \frac{c_{1}g_{1i}+c_{2}g_{2i}}{x_8},~~k=1,\cdots,8
\eea
with $c_1=-c_2$ and taking them into \eref{syz2}, we got two independent recurrence relations
\bea
&&c_{n_1,\cdots,n_7}^{1000000}i[\lam_0;n_1+1,\cdots,n_7]+c_{n_1,\cdots,n_7}^{0010000}i[\lam_0;\cdots,n_3+1,\cdots]+\delta_{11}=0\nn
&&c_{n_1,\cdots,n_7}^{1000000}i[\lam_0;n_1+1,\cdots]+c_{n_1\cdots,n_7}^{0001000}i[\cdots,n_4+1,\cdots]+c_{n_1,\cdots,n_7}^{0000100}i[\cdots,n_5+1,\cdots]\nn&&+c_{n_1,\cdots,n_7}^{0000010}i[\cdots,n_6+1,n_7]+c_{n_1,\cdots,n_7}^{0000001}i[\cdots,n_7+1]+c_{n_1,\cdots,n_7}^{0000000}i[\lam_0;n_1,\cdots,n_7]+\delta_{12}=0  ~~~\Label{step1}
\eea 
Since the two relations could not be used to lower total power, we need to combine them with the generators in degree two and setting $z_i\equiv \frac{\sum_{j=1}^{37} b_j g_{ji} }{ x_8^{2}}$. By choosing proper values of the  coefficients $b_j$, we could get the simplified IBP relations,
\bea
c_{n_1,\cdots,n_7}^{i_1,\cdots,i_7}i[\lam_0;n_1+i_1,\cdots n_7+i_7] +\sum_{\substack{\{i_1,\cdots,i_7\} \in A_0\cup A_9\\ \cup A_{10} \cup A_{11} \cup A_{12} } }  c_{n_1,\cdots,n_7}^{i_1,\cdots,i_7} i[\lam_0;n_1+i_1,\cdots,n_7+i_7] +\delta_{2} = 0,\forall\{i_1,\cdots,i_7\}\in A_5~~~~\Label{topoF22}
\eea

\subsubsection{The example: $I_7^{(0,0)} (2,1,2,1,1,1,1)$}
In this example, we need to reduce $i[\lam_0;1,0,1,0,0,0,0]$. Setting $n_1=\cdots=n_7=0$, we could use one of the relations  \eref{topoF22}, 
\bea
c_{0000000}^{1010000}i[\lam_0;1,0,1,0,0,0,0]+\sum_{j=1,j\neq2}^{7}j^{+}i[\lam_0;0,0,0,0,0,0,0]+c_{0000000}^{0000000}i[\lam_0;0,0,0,0,0,0,0]+\delta_{1,3}&=&0~~~~\Label{step2}~~~~~~~~~
\eea
Note in the relation  \eref{step2} there is no term $i[\lam_0;0,1,0,0,0,0,0]$, but there is one term $i[\lam_0;0,0,1,0,0,0,0]$, which is not our basis. So the next step is to use \eref{step1} to reduce $i[\lam_0;0,0,1,0,0,0,0]$ to $i[\lam_0;1,0,0,0,0,0,0]$.
After the two step, we successfully reduce our target $i[\lam_0;1,0,1,0,0,0,0]$ in the top-sector level. After translating the $i[\lam_0;n_1,\cdots,n_7]$ to $I_7^{(0,0)}(n_1+1,\cdots,n_7+1)$, we obtain the reduction coefficients which are confirmed by FIRE6. In summary,  to get the reduction of the top-sector, we just need to use \eref{step1} and \eref{step2} once respectively.

\subsection{The tensor's case}
For the tensor,  the function  $F(q,y)$ is   
\bea
F(q,y)&=&F(0,0)+\frac{1}{4} y (-2 q_2\cdot  (2 p_2 ((x_1+x_2) x_7-x_3 x_4)-2 p_1 (x_2 x_4+x_3 x_4-x_1 x_7)+2 p_3 (x_3 x_6+x_4 x_6\nn
&&+x_7 x_6+x_3 x_7+x_1 (x_6+x_7)+x_2 (x_6+x_7))+q_1 (x_4+x_7) y)+4 q_1 (p_3 (x_4 x_6-x_5 x_7)\nn
&&+p_1 (x_4 (x_5+x_6)+x_1 (x_4+x_5+x_6+x_7))-p_2 ((x_5+x_6) x_7+x_3 (x_4+x_5+x_6+x_7)))\nn
&&+q_1^2 (x_4+x_5+x_6+x_7) y+q_2^2 (x_1+x_2+x_3+x_4+x_7) y+4 (x_3 x_4+x_5 x_4+x_6 x_4\nn
&&+x_3 x_5+x_3 x_6+x_3 x_7+x_5 x_7+x_6 x_7+x_1 (x_4+x_5+x_6+x_7)+x_2 (x_4+x_5+x_6+x_7)))~~~
\eea
\subsubsection{The  example: $I_7^{(1,0)}(1,1,1,1,1,1,1)$}

By directly calculation and integrate over y, we got 
\bea
&&i[\lam_0,0,\cdots,0]^{\mu}\nn
&=&\frac{-D}{3D-14}\Big( \Big(-(i[\lam_0-1,0,0,0,1,0,1,0]+i[\lam_0-1,0,0,0,1,1,0,0]+i[\lam_0-1,1,0,0,0,0,0,1]\nn&&+i[\lam_0-1,1,0,0,0,0,1,0]+i[\lam_0-1,1,0,0,0,1,0,0]+i[\lam_0-1,1,0,0,1,0,0,0]) p_1^{\mu}\nn&&+(i[\lam_0-1,0,0,0,0,0,1,1]+i[\lam_0-1,0,0,0,0,1,0,1]+i[\lam_0-1,0,0,1,0,0,0,1]+i[\lam_0-1,0,0,1,0,0,1,0]\nn&&+i[\lam_0-1,0,0,1,0,1,0,0]+i[\lam_0-1,0,0,1,1,0,0,0]) p_2^{\mu}+(i[\lam_0-1,0,0,0,0,1,0,1]\nn&&-i[\lam_0-1,0,0,0,1,0,1,0]) p_3^{\mu}\Big)\Big)~~~~~~~~~~~~~~~~~~
\eea
To reduce the term in different dimension, we could solve the syzygy equation
\bea
\sum_{i=1}^8 g_{ji}\frac{\d F}{\d x_i}&=&g_0 \tilde Y
\eea
with 
\bea
\tilde Y&=&x_4 x_5 p_1^{\mu}+x_4 x_6 p_1^{\mu}+x_1 \Big(x_4+x_5+x_6+x_7\Big) p_1^{\mu}-x_5 x_7 p_2^{\mu}-x_6 x_7 p_2^{\mu}-\nn&&x_3 \Big(x_4+x_5+x_6+x_7\Big) p_2^{\mu}+x_4 x_6 p_3^{\mu}-x_5 x_7 p_3^{\mu}
\eea
and then defing the $z_i$s as
\bea
z_i&\equiv &\frac{\sum_{j}c_{j}g_{ji}}{x_8^\alpha}
\eea
Taking the $z_i$s into the IBP identities, we got the results. Since the results are too complicated to write here, we put them in the attached files. Using the similar method, we finally got the reduction results.
\section{Conclusion}

In this article, we considered a new  parametric representation of  Feynman integrals proposed by Chen \cite{chen1,chen2}. In the practical calculation of two-loop integrals, there are large number of IBP relations which makes our calculation hard. To simplify the IBP relations, we used the "syzygy" trick to cancel the dimensional shift and the unwanted doubled propagators in the parametric representation. By this trick, as shown in three examples of two-loop diagrams, we successfully simplified the IBP relations, and used them to reduce the diagrams. The main advantage of our method is that, by the simplified IBP relations, we could directly choose which relation we need to lower the total power, and do the reduction more efficiently than the traditional method, so it will be good if such a  procedure could be programmed.
However, as the number of loops going higher, the homogeneous polynomial  $F$ becomes more and more complicated, which makes the syzygy equations hard to solve. Further more, more mathematical properties and constructions are still need to find and utilize, such as the polynomial tangent space. We will focus on them in the near future.
\section*{Acknowledgments}
I would like to thank Bo Feng for the inspiring guidance.

\newpage
\appendix
\section{The syzygy generators of sunset's case}
In this part, we give the syzygy generators in the section 4.1:

\bean
g_1&=&\Big\{0, 0, s x_3^2,
m_1^4 x_1^2 + m_2^4 x_2^2 + m_3^4 x_3^2 - m_3^2 s x_3^2 +
2 m_3^2 x_3  + x_4^2 + 2 m_2^2 x_2 (m_3^2 x_3 + x_4) +
2 m_1^2 x_1 (m_2^2 x_2 + m_3^2 x_3 + x_4)\nn&&, -m_1^2 x_1 - m_2^2 x_2 -
m_3^2 x_3 - s x_3 - x_4\Big\}\eean
\bean
g_2&=&\Big\{0, 0, -s (m_2^2 + s) x_3^2 + m_1^4 (x_2 x_3 + x_1 (x_2 + x_3)) -
m_1^2 (-2 s x_3^2 + m_2^2 (x_2 x_3 + x_1 (x_2 + x_3))),\nn&&
m_1^6 x_1 (x_1 - x_2) -
m_1^4 (s x_1 (x_1 - x_2) + m_3^2 x_1 x_2 +
m_2^2 x_2 (-4 x_1 + x_2) - 2 m_3^2 x_1 x_3 +
2 m_3^2 x_2 x_3 \nn&&- 3 x_1 x_4 + x_2 x_4) - (m_2^2 +
s) (m_2^4 x_2^2 + m_3^4 x_3^2 + x_4^2 +
2 m_2^2 x_2 (m_3^2 x_3 + x_4) + m_3^2 x_3 (-s x_3 + 2 x_4)) \nn&&+
m_1^2 (m_2^4 x_2 (-x_1 + 3 x_2)+
m_2^2 (-3 s x_1 x_2 + m_3^2 x_2 (x_1 + 6 x_3) - x_1 x_4 +
5 x_2 x_4) +
2 (m_3^4 x_3^2 \nn&&- m_3^2 x_3 (s (x_1 + x_3) - 2 x_4) +
x_4 (-s x_1 + x_4))), -(2 m_1^2 - m_2^2 - s) (m_1^2 x_1 +
m_2^2 x_2 + m_3^2 x_3 + s x_3 + x_4)\Big\}\eean
\bean
g_3&=&\Big\{0, x_2^2, -x_3^2, -m_2^2 x_2^2 + m_3^2 x_3^2, -x_2 + x_3\Big\}\eean
\bean
g_4&=&\Big\{0, 2 m_2^4 x_2^2 - m_1^4 (x_1 + x_2) x_3 +
m_1^2 (-2 m_2^2 x_2^2 + s (x_1 + x_2) x_3),
x_3 (2 (-m_2^4 + s^2) x_3 + m_1^4 (x_1 + x_3)\nn&& +
m_1^2 (2 m_2^2 x_3 - s (x_1 + 3 x_3))),
2 m_1^2 m_2^2 (-m_1^2 + s) x_1 x_2 - 2 (m_2^6 - m_2^4 s) x_2^2 +
m_1^2 (m_1^2 - s) (m_1^2 \nn&&+ m_2^2 - 3 m_3^2 - s) x_1 x_3 +
2 m_2^2 (m_1^2 - 2 m_3^2) (m_1^2 - s) x_2 x_3 -
2 m_3^2 (-m_2^4 + m_1^2 (m_2^2 + m_3^2 - s) + s (-m_3^2 + s)) x_3^2 \nn&&-
2 (m_1^4 - m_1^2 s) x_1 x_4 +
4 m_2^2 (-m_1^2 + s) x_2 x_4 + (m_1^2 - 4 m_3^2) (m_1^2 - s) x_3 x_4 - 2 (m_1^2 - s) x_4^2, \nn&&-2 m_2^2 (-2 m_1^2 + m_2^2 + s) x_2 - (m_1^4 - 2 m_2^4 + m_1^2 (2 m_2^2 - 2 m_3^2 - 3 s) +
2 s (m_3^2 + s)) x_3 + 2 (m_1^2 - s) x_4\Big\}\eean
\bean
g_5&=&\Big\{0, -m_1^2 m_2^2 (m_1^2 - s) (m_1^2 - 2 m_2^2 + s) x_1 x_2 -
m_2^2 (m_1^6 + 2 m_1^4 (m_3^2 - 2 s) +
2 (2 m_2^6 + m_2^4 (2 m_3^2 - 3 s)\nn&& - m_2^2 s^2 + s^2 (-m_3^2 + s)) +
m_1^2 (-6 m_2^4 + (4 m_3^2 - 3 s) s + m_2^2 (-8 m_3^2 + 14 s))) x_2^2 - m_1^2 (m_1^2 - s) (-2 m_2^4 +
m_1^2 (m_2^2 \nn&&+ m_3^2 - s) + (3 m_3^2 - s) s +
m_2^2 (-4 m_3^2 + 3 s)) x_1 x_3 -
m_1^2 (m_1^2 - s) (-2 m_2^4 + m_1^2 (m_2^2 + m_3^2 - s) + (3 m_3^2 - s) s \nn
&&+
m_2^2 (-4 m_3^2 + 3 s)) x_2 x_3,
m_1^2 m_2^2 (m_1^2 - s) (m_1^2 - 2 m_2^2 + s) x_1 x_2 +
m_1^2 (m_1^2 - s) (-2 m_2^4 + m_1^2 (m_2^2 + m_3^2 - s) \nn&&+ (3 m_3^2 - s) s +
m_2^2 (-4 m_3^2 + 3 s)) x_1 x_3 +
m_1^2 m_2^2 (m_1^2 - s) (m_1^2 - 2 m_2^2 + s) x_2 x_3 + (m_1^6 (m_2^2 + m_3^2 - s) \nn&&+ 2 (m_2^2 - s) (2 m_2^6 + m_2^4 (2 m_3^2 - s) + 2 m_2^2 (m_3^2 - s) s +
s^2 (-3 m_3^2 + s)) - 2 m_1^4 (-s^2 + m_2^2 (m_3^2 + s)) \nn&&+
m_1^2 (-6 m_2^6 + m_2^2 (16 m_3^2 - 9 s) s + s^2 (-7 m_3^2 + s) +
m_2^4 (-8 m_3^2 + 14 s))) x_3^2, -m_1^2 m_2^2 (m_1^2 - s) (m_1^4 + 2 m_2^4 \nn&&+
m_1^2 (-3 m_2^2 + 3 m_3^2 - 2 s) + (7 m_3^2 - 3 s) s +
5 m_2^2 (-2 m_3^2 + s)) x_1 x_2 +
2 m_2^4 (m_2^2 - s) (m_1^4 \nn
&&+ 2 m_2^4 + m_2^2 (2 m_3^2 - s) - 2 m_3^2 s +
m_1^2 (-3 m_2^2 + s)) x_2^2 +
m_1^2 (m_1^2 - s) (-2 m_2^6 + m_1^4 (m_2^2 + m_3^2 - s) - 9 m_3^4 s \nn&&+ s^3 +
m_2^4 (-2 m_3^2 + 5 s) -
m_1^2 (m_2^4 + 3 m_3^4 + m_2^2 (4 m_3^2 - s) - 5 m_3^2 s) +
4 m_2^2 (3 m_3^4 - s^2)) x_1 x_3 +
2 m_2^2 (m_1^6 (m_2^2 - s) \nn&&-
2 m_1^4 (m_2^4 + m_3^4 + m_2^2 (m_3^2 - s) - 2 m_3^2 s) +
2 m_3^2 s ((3 m_3^2 - s) s + m_2^2 (-4 m_3^2 + 2 s)) \nn&&+
m_1^2 (2 m_2^4 s + m_2^2 (8 m_3^4 - 2 m_3^2 s - 3 s^2) +
s (-4 m_3^4 - 2 m_3^2 s + s^2))) x_2 x_3 -
2 m_3^2 (2 m_2^8 + m_1^4 (m_2^4 + m_2^2 (m_3^2 - 2 s) \nn&&+ (m_3^2 - s)^2) +
m_2^6 (2 m_3^2 - 3 s) - m_2^4 s^2 - s^2 (3 m_3^4 - 4 m_3^2 s + s^2) +
m_2^2 s (4 m_3^4 - 7 m_3^2 s + 3 s^2) \nn&&-
m_1^2 (3 m_2^6 + m_2^4 (4 m_3^2 - 6 s) + 2 m_3^2 s (-m_3^2 + s) +
m_2^2 (4 m_3^4 - 8 m_3^2 s + 3 s^2))) x_3^2 +
2 m_1^2 (m_1^2 - s) (m_2^2 (4 m_3^2 - 2 s) \nn&&+ s (-3 m_3^2 + s) +
m_1^2 (-m_3^2 + s)) x_1 x_4 -
m_2^2 (m_1^2 - s) (m_1^4 + m_1^2 (-2 m_2^2 + 4 m_3^2 - 3 s) -
4 (m_2^2 (4 m_3^2 - 2 s) \nn&&+ s (-3 m_3^2 + s))) x_2 x_4 + (m_1^6 (m_2^2 + m_3^2 - s) -
2 m_1^4 (m_2^4 + 2 m_3^4 + m_2^2 (2 m_3^2 - s) - 3 m_3^2 s) +
4 m_3^2 s ((3 m_3^2 - s) s \nn&&+ m_2^2 (-4 m_3^2 + 2 s)) +
m_1^2 (2 m_2^4 s + m_2^2 (16 m_3^4 - 4 m_3^2 s - 3 s^2) +
s (-8 m_3^4 - 3 m_3^2 s + s^2))) x_3 x_4 \nn&&+
2 (m_1^2 - s) (m_2^2 (4 m_3^2 - 2 s) + s (-3 m_3^2 + s) +
m_1^2 (-m_3^2 + s)) x_4^2,
m_2^2 (m_1^6 + 4 m_2^6 + m_1^4 (4 m_3^2 - 6 s) \nn
&&+ m_2^4 (4 m_3^2 - 6 s) +
2 m_2^2 (4 m_3^2 - 3 s) s + 4 s^2 (-2 m_3^2 + s) +
m_1^2 (-6 m_2^4 - 2 m_2^2 (8 m_3^2 - 9 s) + (8 m_3^2 - 3 s) s)) x_2 \nn&&+ (-m_1^6 (m_2^2 + m_3^2 - s) +
2 m_1^4 (m_3^4 - m_3^2 s - s^2 + m_2^2 (m_3^2 + s)) +
2 (-2 m_2^8 + m_2^4 s^2 + m_2^6 (-2 m_3^2 + 3 s)\nn&& +
m_2^2 s (4 m_3^4 + 3 m_3^2 s - 3 s^2) +
s^2 (-3 m_3^4 - 2 m_3^2 s + s^2)) +
m_1^2 (6 m_2^6 + 2 m_2^4 (4 m_3^2 - 7 s) \nn&&+
s (4 m_3^4 + 7 m_3^2 s - s^2) +
m_2^2 (-8 m_3^4 - 12 m_3^2 s + 9 s^2))) x_3 -
2 (m_1^2 - s) (m_2^2 (4 m_3^2 - 2 s)\nn&& + s (-3 m_3^2 + s) +
m_1^2 (-m_3^2 + s)) x_4\Big\}\nn
g_6&=&\Big\{0, -m_1^4 x_1 (x_1 + x_2) + m_2^4 x_2 (x_1 + x_2) +
m_2^2 (-s x_2 (3 x_1 + x_2) +
m_3^2 (x_2 x_3 + x_1 (2 x_2 + x_3))) + s (x_1 + x_2) x_4 \nn&&-
m_1^2 (-s x_1 (x_1 + 3 x_2) +
m_3^2 (x_2 x_3 + x_1 (2 x_2 + x_3)) + (x_1 + x_2) x_4),
m_1^4 x_1 (x_1 + x_3) + m_2^4 (x_2 x_3 + x_1 (x_2 + 2 x_3)) \nn&&+
m_2^2 (m_3^2 x_3 (x_1 + x_3) -
s (x_2 x_3 + x_1 (x_2 + 4 x_3))) -
m_1^2 (x_1 + x_3) (s x_1 + m_3^2 x_3 - x_4) +
s (2 s x_1 x_3 - (x_1 + x_3) x_4),\nn&& -m_1^2 (m_1^2 - s) (m_1^2 -
5 m_2^2 + 3 m_3^2 + s) x_1^2 -
m_2^2 (m_2^2 - s) (-5 m_1^2 + m_2^2 + 3 m_3^2 + s) x_1 x_2 +
m_3^2 (-3 m_1^4 - 3 m_2^4 \nn&&+ 2 (m_3^2 - s) s + m_2^2 (-m_3^2 + s) +
m_1^2 (6 m_2^2 - m_3^2 + s)) x_1 x[
3] + (-2 m_1^4 + m_1^2 (7 m_2^2 - 3 m_3^2 - s) +
s (-5 m_2^2 \nn&&+ 3 m_3^2 + s)) x_1 x_4 +
3 (m_2^4 - m_2^2 s) x_2 x_4 +
3 (-m_1^2 + m_2^2) m_3^2 x_3 x_4 - (m_1^2 - 2 m_2^2 + s) x_4^2, (m_1^4 - m_1^2 (4 m_2^2 - 4 m_3^2 \nn&&+ s) -
2 (m_2^4 + m_2^2 (m_3^2 - 4 s) + s (m_3^2 + s))) x_1 -
3 (m_2^4 - m_2^2 s) x_2 +
3 (m_1^2 - m_2^2) m_3^2 x_3 + (m_1^2 - 2 m_2^2 + s) x_4\Big\}\nonumber
\eea
\bea
g_7&=&\Big\{x_3 (m_3^2 x_3 + x_4),
2 s x_2 x_3 - m_1^2 (x_1 + x_2) x_3 -
m_3^2 x_3 (2 x_2 + x_3) - 2 x_2 x_4 - x_3 x_4,
x_3 (-m_3^2 x_3 + m_1^2 (x_1 + x_3) - x_4),\nn&&
m_1^2 (-m_1^2 + 3 m_2^2 - 3 m_3^2 + s) x_1 x_3 +
2 (m_2^4 - m_2^2 s) x_2 x_3 +
m_3^2 (-3 m_1^2 + 3 m_2^2 - m_3^2 + s) x_3^2 -
2 m_1^2 x_1 x_4 + (-2 m_1^2 \nn&&+ 3 m_2^2 - 3 m_3^2 + s) x_3 x_4 -
2 x_4^2, (m_1^2 - 2 (m_2^2 - 2 m_3^2 + s)) x_3 + 4 x_4\Big\}\nn
g_8&=&\Big\{x_3 (-m_2^4 x_2 + m_2^2 (s x_2 - m_3^2 x_3 - x_4) +
m_1^2 (m_3^2 x_3 + x_4)), -m_1^4 (x_1 + x_2) x_3 +
m_2^2 (m_2^2 x_2 x_3 - 3 s x_2 x_3 \nn&&+ m_3^2 x_3 (2 x_2 + x_3) +
2 x_2 x_4 + x_3 x_4) +
m_1^2 (s (x_1 + 3 x_2) x_3 - (2 x_2 + x_3) (m_3^2 x_3 + x_4)),
\nn&&x_3 (2 s^2 x_3 + m_1^4 (x_1 + x_3) + m_2^4 (x_2 + 2 x_3) -
m_1^2 (m_3^2 x_3 + s (x_1 + x_3) + x_4) +
m_2^2 (m_3^2 x_3 - s (x_2 + 4 x_3) + x_4)),\nn&& -m_1^2 (m_1^2 -
s) (m_1^2 - 5 m_2^2 + 3 m_3^2 + s) x_1 x_3 -
m_2^2 (m_2^2 - s) (-5 m_1^2 + m_2^2 + 3 m_3^2 + s) x_2 x_3 +
m_3^2 (-3 m_1^4 - 3 m_2^4 \nn&&+ 2 (m_3^2 - s) s + m_2^2 (-m_3^2 + s) +
m_1^2 (6 m_2^2 - m_3^2 + s)) x_3^2 - 2 (m_1^4 - m_1^2 s) x_1 x_4 -
2 (m_2^4 - m_2^2 s) x_2 x_4 \nn&&- (2 m_1^4 + m_2^4 + m_1^2 (-6 m_2^2 + 3 m_3^2) - 4 m_3^2 s +
m_2^2 (m_3^2 + 3 s)) x_3 x_4 -
2 (m_1^2 - s) x_4^2, (m_1^4 - m_1^2 (4 m_2^2 - 4 m_3^2 + s) \nn&&-
2 (m_2^4 + m_2^2 (m_3^2 - 4 s) + s (m_3^2 + s))) x_3 + (4 m_1^2 - 2 (m_2^2 + s)) x_4\Big\}\nn
g_9&=&\Big\{m_2^4 x_2 (-x_2 + x_3) +
m_2^2 (x_2 - x_3) (s x_2 - m_3^2 x_3 - x_4) + s x_2 x_4 -
m_1^2 x_3 (m_3^2 (-x_2 + x_3) + x_4),
m_2^4 x_2 (x_2 - x_3) - m_1^4 (x_1 + x_2) (x_2 - x_3) \nn&&+
m_1^2 (s (x_1 + 3 x_2) (x_2 - x_3) - (2 x_2 +
x_3) (m_3^2 (x_2 - x_3) - x_4)) - s x_2 x_4 -
m_2^2 (3 s x_2 (x_2 - x_3) +
m_3^2 (-2 x_2^2 \nn&&+ x_2 x_3 + x_3^2) + (x_2 + x_3) x_4),
m_1^4 (x_2 - x_3) (x_1 + x_3) +
m_2^4 (x_2^2 + x_2 x_3 - 2 x_3^2) +
m_2^2 (m_3^2 (x_2 - x_3) x_3
\eean
\bean
&&-
s (x_2^2 + 3 x_2 x_3 - 4 x_3^2) + (x_2 + x_3) x_4) +
s (2 s (x_2 - x_3) x_3 - (x_2 + 2 x_3) x_4) +
m_1^2 (-s (x_2 - x_3) (x_1 + x_3) \nn&&+
x_3 (m_3^2 (-x_2 + x_3) + x_4)), -m_1^2 (m_1^2 - s) (m_1^2 -
5 m_2^2 + 3 m_3^2 + s) x_1 x_2 -
m_2^2 (m_2^2 - s) (-5 m_1^2 \nn&&+ m_2^2 + 3 m_3^2 + s) x_2^2 +
m_1^2 (m_1^2 - s) (m_1^2 - 5 m_2^2 + 3 m_3^2 + s) x_1 x_3 - (-m_2^6 + 3 m_1^4 m_3^2 \nn&&+ 2 m_3^2 s (-m_3^2 + s) +
m_2^2 (m_3^2 + s)^2 +
m_1^2 (5 m_2^4 + m_3^4 - m_3^2 s - m_2^2 (6 m_3^2 + 5 s))) x_2 x_3 \nn&&+
m_3^2 (3 m_1^4 + 3 m_2^4 + m_2^2 (m_3^2 - s) +
m_1^2 (-6 m_2^2 + m_3^2 - s) + 2 s (-m_3^2 + s)) x_3^2 +
4 (m_1^4 - m_1^2 s) x_1 x_4 \nn&&+ (-m_1^4 + m_2^4 + s (3 m_3^2 + s) - m_2^2 (m_3^2 + 6 s) +
m_1^2 (7 m_2^2 - 2 (m_3^2 + s))) x_2 x_4 + (2 m_1^4 + m_2^4 + m_1^2 (-6 m_2^2 + 5 m_3^2) \nn&&-
m_2^2 (m_3^2 - 3 s) - 4 m_3^2 s) x_3 x_4 +
4 (m_1^2 - s) x_4^2, (m_1^4 - m_1^2 (4 m_2^2 - 4 m_3^2 + s) -
2 (m_2^4 + m_2^2 (m_3^2 - 4 s) + s (m_3^2 + s))) x_2 \nn&&+ (-m_1^4 + m_1^2 (4 m_2^2 - 4 m_3^2 + s) +
2 (m_2^4 + m_2^2 (m_3^2 - 4 s) + s (m_3^2 + s))) x_3 -
6 (m_1^2 - s) x_4\Big\}\nn
g_{10}&=&\Big\{x_1, x_2, x_3, x_4, -3\Big\}
\eean
\section{The coefficients of sunset}
\subsection{The analytical coefficients of sunset example: $I_3^{(0,0)}(3,1,1)$}
The reduction result is
\bea
I_3^{(0,0)}(3,1,1)&=&c_{311\to211}I_3^{(0,0)}(2,1,1)+c_{311\to121}I_3^{(0,0)}(1,2,1)+c_{311\to112}I_3^{(0,0)}(1,1,2)+c_{311\to111}I_3^{(0,0)}(1,1,1)+\cdots~~~~
\eea
with the coefficients
\bea
c_{311\to211}&=&\frac{n_{311\to211}}{N_{311\to211}},~~c_{311\to121}=\frac{n_{311\to121}}{N_{311\to121}},~~c_{311\to112}=\frac{n_{311\to112}}{N_{311\to112}},~~c_{311\to111}=\frac{n_{311\to111}}{N_{311\to111}}
\eea
where
\bean
n_{311\to211}&=&2 m_1^4 (2 m_2^2 ((7 D-22) m_3^2+(D-4) s)+(5 D-18) m_2^4+2 (D-4) m_3^2 s+(5 D-18) m_3^4+3 (3 D-10) s^2)\nn&&-4 m_1^2 (m_2^2 (2 (8 D-29) m_3^2 s-(D-4) m_3^4+(10-3 D) s^2)+m_2^4 (s-(D-4) m_3^2)+(D-4) m_2^6\nn&&+(m_3^2-s)^2 ((D-4) m_3^2+(2 D-7) s))-4 m_1^6 ((3 D-10) m_2^2+(3 D-10) m_3^2+(4 D-13) s)\nn&&+(5 D-16) m_1^8+(D-4) (-2 m_2^2 (m_3^2+s)+m_2^4+(m_3^2-s)^2)^2\nn
N_{311\to211}&=&4 m_1^2 (m_1^2+2 m_1 (m_2-m_3)+m_2^2-2 m_2 m_3+m_3^2-s) (m_1^2-2 m_1 m_2+2 m_1 m_3\nn&&+m_2^2-2 m_2 m_3+m_3^2-s) (m_1^2-2 m_1 (m_2+m_3)+m_2^2+2 m_2 m_3+m_3^2-s) (m_1^2+2 m_1 (m_2+m_3)\nn&&+m_2^2+2 m_2 m_3+m_3^2-s)\nn
\eean
\bea
n_{311\to121}&=&(D-3) m_2^2 (-2 m_1^2 (m_2^2-s) (m_2^2-3 m_3^2+s)+m_1^4 (m_2^2-s)+m_2^2 (m_3^4+3 s^2)\nn&&-m_2^4 (2 m_3^2+3 s)+m_2^6-s (m_3^2-s)^2)\nn
N_{311\to121}&=&m_1^2 (m_1^2+2 m_1 (m_2-m_3)+m_2^2-2 m_2 m_3+m_3^2-s) (m_1^2-2 m_1 m_2+2 m_1 m_3\nn&&+m_2^2-2 m_2 m_3+m_3^2-s) (m_1^2-2 m_1 (m_2+m_3)+m_2^2+2 m_2 m_3+m_3^2-s) (m_1^2+2 m_1 (m_2+m_3)\nn&&+m_2^2+2 m_2 m_3+m_3^2-s)\nn
n_{311\to112}&=&(D-3) m_3^2 (2 m_1^2 (m_3^2-s) (3 m_2^2-m_3^2-s)+m_1^4 (m_3^2-s)-2 m_2^2 (m_3^4-s^2)+m_2^4 (m_3^2-s)+(m_3^2-s)^3)\nn
N_{311\to112}&=&m_1^2 (m_1^2+2 m_1 (m_2-m_3)+m_2^2-2 m_2 m_3+m_3^2-s) (m_1^2-2 m_1 m_2+2 m_1 m_3\nn&&+m_2^2-2 m_2 m_3+m_3^2-s) (m_1^2-2 m_1 (m_2+m_3)+m_2^2+2 m_2 m_3+m_3^2-s) (m_1^2+2 m_1 (m_2+m_3)+m_2^2\nn&&+2 m_2 m_3+m_3^2-s)\nn
n_{311\to111}&=&(3-D) (3 d-8) (-m_1^2 (2 m_2^2 (s-5 m_3^2)+m_2^4+2 m_3^2 s+m_3^4-3 s^2)\nn&&-m_1^4 (m_2^2+m_3^2+3 s)+m_1^6+((m_2-m_3)^2-s) (m_2^2+m_3^2-s) ((m_2+m_3)^2-s))\nn
N_{311\to111}&=&4 m_1^2 ((m_1+m_2-m_3)^2-s) ((m_1-m_2\nn&&+m_3)^2-s) ((-m_1+m_2+m_3)^2-s) ((m_1+m_2+m_3)^2-s)
\eea
\\
\\
\\
\subsection{The analytical coefficients of sunset example: $I_3^{(0,0)}(2,2,1)$}
The reduction result is
\bea
I_3^{(0,0)}(2,2,1)&=&c_{221\to211}I_3^{(0,0)}(2,1,1)+c_{221\to121}I_3^{(0,0)}(1,2,1)+c_{221\to112}I_3^{(0,0)}(1,1,2)+c_{221\to111}I_3^{(0,0)}(1,1,1)+\cdots ~~~~~
\eea
with the coefficients
\bea
c_{221\to211}&=&\frac{n_{221\to211}}{N_{221\to211}},~~c_{211\to121}=\frac{n_{221\to121}}{N_{221\to121}},~~c_{221to112}=\frac{n_{221\to112}}{N_{221\to121}},~c_{221\to111}=\frac{n_{221\to111}}{N_{221\to111}}
\eea
where
\bea
n_{221\to211}&=&(D-3) (-m_1^2 (2 m_2^2 (m_3^2-3 s)+5 m_2^4+14 m_3^2 s-7 m_3^4+s^2)+m_1^4 (7 m_2^2-3 m_3^2+5 s)\nn&&-3 m_1^6+m_2^2 (2 m_3^2 s+3 m_3^4+3 s^2)-3 m_2^4 (m_3^2+s)+m_2^6-(m_3^2-s)^2 (m_3^2+s))\nn
N_{221\to211}&=&((m_1+m_2-m_3)^2-s) ((m_1-m_2+m_3)^2-s) ((-m_1+m_2+m_3)^2-s) ((m_1+m_2+m_3)^2-s)\nn
n_{211\to121}&=&(D-3) (m_1^2 (-2 m_2^2 (m_3^2-3 s)+7 m_2^4+2 m_3^2 s+3 m_3^4+3 s^2)-m_1^4 (5 m_2^2+3 (m_3^2+s))\nn&&+m_1^6+m_2^2 (-14 m_3^2 s+7 m_3^4-s^2)+m_2^4 (5 s-3 m_3^2)-3 m_2^6-(m_3^2-s)^2 (m_3^2+s))\nn
N_{211\to121}&=&((m_1+m_2-m_3)^2-s) ((m_1-m_2+m_3)^2-s) ((-m_1+m_2+m_3)^2-s) ((m_1+m_2+m_3)^2-s)\nn
n_{221\to112}&=&8 (D-3) m_3^2 (m_3^2-s) (-m_1^2-m_2^2+m_3^2+s)\nn
N_{221\to112}&=&((m_1+m_2-m_3)^2-s) ((m_1-m_2+m_3)^2-s) ((-m_1+m_2+m_3)^2-s) ((m_1+m_2+m_3)^2-s)\nn
n_{221\to111}&=&(D-3) (3 D-8) (-2 m_1^2 (m_2^2-m_3^2+s)+m_1^4+2 m_2^2 (m_3^2-s)+m_2^4+2 m_3^2 s-3 m_3^4+s^2)\nn
N_{221\to111}&=&(m_1^2+2 m_1 (m_2-m_3)+m_2^2-2 m_2 m_3+m_3^2-s) (m_1^2-2 m_1 m_2+2 m_1 m_3+m_2^2-2 m_2 m_3+m_3^2\nn&&-s) (m_1^2-2 m_1 (m_2+m_3)+m_2^2+2 m_2 m_3+m_3^2-s) (m_1^2+2 m_1 (m_2+m_3)+m_2^2+2 m_2 m_3+m_3^2-s)~~~
\eea
\\
\\
\\
\\
\\
\\
\\
\\
\\
\\
\\
\\
\\
\\
\\
\\
\subsection{The coefficients of $i(\lam_0-2;0,2,2)$}
\bea
c_{i;022\to000}&=&\frac{4 (3 D-1)}{3 (D+2) s (2 (m_{2}^2+s)-m_{1}^2)}\nn
c_{i;022\to100}&=&\frac{4 (1-3 D) m_{1}^2}{3 (D+2) s (m_{1}^2-2 (m_{2}^2+s))}\nn
c_{i;022\to010}&=&\frac{1}{3 (D+2) s (2 (m_{2}^2+s)-m_{1}^2) (m_{1}^2 (2 m_{2}^2-3 s)+m_{1}^4-2
	(m_{2}^2-s) (m_{2}^2-m_{3}^2+s))}\times \Big\{D (6 m_{1}^2 (m_{2}^2 (3 m_{3}^2\nn&&-23 s)+8 m_{2}^4-3 s (m_{3}^2-3 s))+45 m_{1}^4 m_{2}^2-9 m_{1}^6-12 (-8 m_{2}^2 s+5
m_{2}^4+3 s^2) (m_{2}^2-m_{3}^2+s))\nn&&-2 (m_{2}^2-s) (3 m_{1}^2+2 m_{2}^2-6 s) (5 m_{1}^2-4
(m_{2}^2-m_{3}^2+s))\Big\}\nn
c_{i;022\to001}&=&\frac{1}{3 (D+2) s
	(2 (m_{2}^2+s)-m_{1}^2) (m_{1}^2 (2 m_{2}^2-3 s)+m_{1}^4-2 (m_{2}^2-s) (m_{2}^2-m_{3}^2+s))}\times\Big\{m_{1}^4 ((22-27 D) m_{2}^2\nn&&+2 (15 D-8) m_{3}^2+2 (11-6 D) s)+2 m_{1}^2 (m_{2}^2 ((15 D-8) m_{3}^2-6 D s)+2 (6 D+1) m_{2}^4\nn&&+2 s ((8-15
D) m_{3}^2+(6 D-11) s))+(3 D-4) m_{1}^6-4 (m_{2}^2-s) (m_{2}^2-m_{3}^2+s) ((15 D-8) m_{3}^2\nn&&-6 D s+4 m_{2}^2+8 s)\Big\}\nn
c_{i;022\to110}&=&\frac{(3 D-2) m_{1}^2 (3 m_{1}^2+2 m_{2}^2-6 s) (m_{1}^2 (2 m_{3}^2-5 m_{2}^2)+m_{1}^4+(3 m_{2}^2-s)
	(m_{2}^2-m_{3}^2+s))}{3 (D+2) s (m_{1}^2-2 (m_{2}^2+s)) (m_{1}^2 (2 m_{2}^2-3 s)+m_{1}^4-2
	(m_{2}^2-s) (m_{2}^2-m_{3}^2+s))}\nn
c_{i;022\to101}&=&-\frac{1}{3 (D+2)
	s (m_{1}^2-2 (m_{2}^2+s)) (m_{1}^2 (2 m_{2}^2-3 s)+m_{1}^4-2 (m_{2}^2-s) (m_{2}^2-m_{3}^2+s))}\times \Big\{(3 D\nn&&-2) m_{1}^2 (m_{1}^2 (5 m_{2}^2 m_{3}^2-3 m_{2}^4+3 s (m_{3}^2+5 s))-3 m_{1}^4 (m_{2}^2+4 s)+3 m_{1}^6+2
(m_{2}^2 (2 m_{3}^2 s\nn&&+6 m_{3}^4-3 s^2)+m_{2}^4 (3 s-9 m_{3}^2)+3 m_{2}^6-3 s (-m_{3}^2 s+2 m_{3}^4+s^2)))\Big\}\nonumber
\eea
\bea
c_{i;022\to011}&=&\frac{1}{3 (D+2) s (2 (m_{2}^2+s)-m_{1}^2) (m_{1}^2 (2 m_{2}^2-3 s)+m_{1}^4-2 (m_{2}^2-s)
	(m_{2}^2-m_{3}^2+s))}\times \Big\{(3 D-2) (2 m_{1}^2 (m_{2}^2 \nn&&(-23 m_{3}^2 s+3 m_{3}^4+13 s^2)+m_{2}^4 (4 m_{3}^2+6 s)+m_{2}^6-3 s (-3 m_{3}^2 s+m_{3}^4+4
s^2))+m_{1}^6 (7 m_{2}^2-3 (3 m_{3}^2+s))\nn&&+m_{1}^4 (m_{2}^2 (14 m_{3}^2-25 s)-10 m_{2}^4+15 s
(m_{3}^2+s))+4 (m_{2}^2-s) (m_{2}^2 (-4 m_{3}^2 s+5 m_{3}^4-s^2)+m_{2}^4 (3 s-6 m_{3}^2)\nn&&+m_{2}^6-3 s
(m_{3}^2-s)^2))\Big\}\nn
c_{i;022\to200}&=&0\nn
c_{i;022\to020}&=&\frac{(3 D-2) m_{2}^2 (m_{2}^2-s) (3 m_{1}^2+2 m_{2}^2-6 s) (5 m_{1}^2-4 (m_{2}^2-m_{3}^2+s))}{3 (D+2) s (2
	(m_{2}^2+s)-m_{1}^2) (m_{1}^2 (2 m_{2}^2-3 s)+m_{1}^4-2 (m_{2}^2-s) (m_{2}^2-m_{3}^2+s))}\nn
c_{i;022\to002}&=&-\frac{1}{3 (D+2) s (2
	(m_{2}^2+s)-m_{1}^2) (m_{1}^2 (2 m_{2}^2-3 s)+m_{1}^4-2 (m_{2}^2-s) (m_{2}^2-m_{3}^2+s))}\times \Big\{(3 D-2) m_{3}^2 (m_{1}^4 \nn&&(13 m_{2}^2-6 m_{3}^2+3 s)+m_{1}^2 (2 m_{2}^2 s-6 m_{2}^4+6 s (m_{3}^2-2 s))\nn&&-4 (m_{2}^2
(-2 m_{3}^2 s+3 m_{3}^4-s^2)+m_{2}^4 (3 s-4 m_{3}^2)+m_{2}^6-3 s (m_{3}^2-s)^2))\Big\}\nonumber
\eea
\nocite{}
\bibliographystyle{JHEP}

\bibliography{reference2}%

\providecommand{\href}[2]{#2}\begingroup\raggedright\begin{thebibliography}{10}

\bibitem{chen1}
W.~Chen, \emph{{{Reduction of Feynman Integrals in the Parametric
  Representation}}}, \href{https://doi.org/10.1007/JHEP02(2020)115}{\emph{JHEP}
  {\bfseries 02} (2020) 115}
  [\href{https://arxiv.org/abs/1902.10387}{{\ttfamily 1902.10387}}].

\bibitem{chen2}
W.~Chen, \emph{{Reduction of Feynman Integrals in the Parametric Representation
  II: Reduction of Tensor Integrals}},
  \href{https://arxiv.org/abs/arXiv:1912.08606}{{\ttfamily arXiv:1912.08606}}.

\bibitem{Xu:2018eos}
X.~Xu and L.L.~Yang, \emph{{Towards a new approximation for pair-production and
  associated-production of the Higgs boson}},
  \href{https://doi.org/10.1007/JHEP01(2019)211}{\emph{JHEP} {\bfseries 01}
  (2019) 211} [\href{https://arxiv.org/abs/1810.12002}{{\ttfamily
  1810.12002}}].

\bibitem{smir}
V.A.~Smirnov, \emph{Evaluating feynman integrals}, {\emph{Springer Tracts in
  Modern Physics} {\bfseries 211} (2005) }.

\bibitem{ibp1}
F.V.~Tkachov, \emph{{A Theorem on Analytical Calculability of Four Loop
  Renormalization Group Functions}},
  \href{https://doi.org/10.1016/0370-2693(81)90288-4}{\emph{Phys. Lett. B}
  {\bfseries 100} (1981) 65}.

\bibitem{ibp2}
K.G.~Chetyrkin and F.V.~Tkachov, \emph{{Integration by Parts: The Algorithm to
  Calculate beta Functions in 4 Loops}},
  \href{https://doi.org/10.1016/0550-3213(81)90199-1}{\emph{Nucl. Phys. B}
  {\bfseries 192} (1981) 159}.

\bibitem{Bern:1992em}
Z.~Bern, L.J.~Dixon and D.A.~Kosower, \emph{{Dimensionally regulated one loop
  integrals}}, \href{https://doi.org/10.1016/0370-2693(93)90400-C}{\emph{Phys.
  Lett. B} {\bfseries 302} (1993) 299}
  [\href{https://arxiv.org/abs/hep-ph/9212308}{{\ttfamily hep-ph/9212308}}].

\bibitem{Bern:1993kr}
Z.~Bern, L.J.~Dixon and D.A.~Kosower, \emph{{Dimensionally regulated pentagon
  integrals}}, \href{https://doi.org/10.1016/0550-3213(94)90398-0}{\emph{Nucl.
  Phys. B} {\bfseries 412} (1994) 751}
  [\href{https://arxiv.org/abs/hep-ph/9306240}{{\ttfamily hep-ph/9306240}}].

\bibitem{Baikov:1996rk}
P.A.~Baikov, \emph{{Explicit solutions of the three loop vacuum integral
  recurrence relations}},
  \href{https://doi.org/10.1016/0370-2693(96)00835-0}{\emph{Phys. Lett. B}
  {\bfseries 385} (1996) 404}
  [\href{https://arxiv.org/abs/hep-ph/9603267}{{\ttfamily hep-ph/9603267}}].

\bibitem{Baikov:1996cd}
P.A.~Baikov, \emph{{Explicit solutions of n loop vacuum integral recurrence
  relations}},  \href{https://arxiv.org/abs/hep-ph/9604254}{{\ttfamily
  hep-ph/9604254}}.

\bibitem{Kosower}
J.~Gluza, K.~Kajda and D.A.~Kosower, \emph{Towards a basis for planar two-loop
  integrals},  \href{https://arxiv.org/abs/arXiv:1009.0472}{{\ttfamily
  arXiv:1009.0472}}.

\bibitem{zhang1}
K.J.~Larsen and Y.~Zhang, \emph{Integration-by-parts reductions from unitarity
  cuts and algebraic geometry},
  \href{https://arxiv.org/abs/arXiv:1511.01071}{{\ttfamily arXiv:1511.01071}}.

\bibitem{zhang2}
K.J.~Larsen and Y.~Zhang, \emph{Integration-by-parts reductions from the
  viewpoint of computational algebraic geometry},
  \href{https://arxiv.org/abs/arXiv:1606.09447}{{\ttfamily arXiv:1606.09447}}.

\bibitem{Larsen:2015ped}
K.J.~Larsen and Y.~Zhang, \emph{{Integration-by-parts reductions from unitarity
  cuts and algebraic geometry}},
  \href{https://doi.org/10.1103/PhysRevD.93.041701}{\emph{Phys. Rev. D}
  {\bfseries 93} (2016) 041701}
  [\href{https://arxiv.org/abs/1511.01071}{{\ttfamily 1511.01071}}].

\bibitem{Larsen:2016tdk}
K.J.~Larsen and Y.~Zhang, \emph{{Integration-by-parts reductions from the
  viewpoint of computational algebraic geometry}},
  \href{https://doi.org/10.22323/1.260.0029}{\emph{PoS} {\bfseries LL2016}
  (2016) 029} [\href{https://arxiv.org/abs/1606.09447}{{\ttfamily
  1606.09447}}].

\bibitem{Zhang:2016kfo}
Y.~Zhang, \emph{{Lecture Notes on Multi-loop Integral Reduction and Applied
  Algebraic Geometry}},  12, 2016
  [\href{https://arxiv.org/abs/1612.02249}{{\ttfamily 1612.02249}}].

\bibitem{Jiang:2017phk}
Y.~Jiang and Y.~Zhang, \emph{{Algebraic geometry and Bethe ansatz. Part I. The
  quotient ring for BAE}},
  \href{https://doi.org/10.1007/JHEP03(2018)087}{\emph{JHEP} {\bfseries 03}
  (2018) 087} [\href{https://arxiv.org/abs/1710.04693}{{\ttfamily
  1710.04693}}].

\bibitem{2021General}
H.~Wang, \emph{General one-loop reduction in generalized feynman
  parametrization form}, .

\bibitem{Smirnov:2019qkx}
A.V.~Smirnov and F.S.~Chuharev, \emph{{FIRE6: Feynman Integral REduction with
  Modular Arithmetic}},
  \href{https://doi.org/10.1016/j.cpc.2019.106877}{\emph{Comput. Phys. Commun.}
  {\bfseries 247} (2020) 106877}
  [\href{https://arxiv.org/abs/1901.07808}{{\ttfamily 1901.07808}}].

\end{thebibliography}\endgroup
\end{document}